\documentclass{aa}
\usepackage{txfonts}
\usepackage{graphicx}
\usepackage{natbib}
\usepackage{mathptm}
\usepackage{psfig}

\begin{document}

\title{A catalogue of the Chandra Deep Field South with multi-colour
   classification and photometric redshifts from COMBO-17}

\author{C. Wolf\inst{1} \and K. Meisenheimer\inst{2} 
   \and M. Kleinheinrich\inst{2} \and A. Borch\inst{2} \and S. Dye\inst{3}
   \and M. Gray\inst{4} \and L. Wisotzki\inst{5} 
   \and E. F. Bell\inst{2} \and H.-W. Rix\inst{2} 
   \and A. Cimatti\inst{6} \and G. Hasinger\inst{7} \and G. Szokoly\inst{7}
}

\institute{ Department of Physics, Denys Wilkinson Bldg.,
            University of Oxford, Keble Road, Oxford, OX1 3RH, U.K. 
       \and Max-Planck-Institut f\"ur Astronomie, K\"onigstuhl 17,
            D-69117 Heidelberg, Germany 
       \and Astrophysics Group, Blackett Lab,
            Imperial College, Prince Consort Road, London, U.K.  
       \and School of Physics and Astronomy,
            University of Nottingham, Nottingham, NG7 2RD, U.K.  
       \and Astrophysikalisches Institut Potsdam, 
            An der Sternwarte 16, D-14482 Potsdam, Germany
       \and Istituto Nazionale di Astrofisica (INAF), Osservatorio 
            Astrofisico di Arcetri, Largo E. Fermi 5, I-50125 Firenze, Italy
       \and Max-Planck-Institut f\"ur Extraterrestrische Physik,
            Postfach 1312, D-85741 Garching, Germany
}

\date{Received / Accepted }

\abstract{
We present the COMBO-17 object catalogue of the Chandra Deep Field South for 
public use, covering a field which is $31\farcm5 \times 30\arcmin$ in size. 
This catalogue lists astrometry, photometry in 17 passbands from 350 to 
930~nm, and ground-based morphological data for 63,501 objects. The catalogue 
also contains multi-colour classification into the categories {\it Star}, 
{\it Galaxy} and {\it Quasar} as well as photometric redshifts. We include 
restframe luminosities in Johnson, SDSS and Bessell passbands and estimated 
errors. The redshifts are most reliable at $R<24$, where the sample contains
approximately 100 quasars, 1000 stars and 10000 galaxies. We use nearly 
1000 spectroscopically identified objects in conjunction with detailed 
simulations to characterize the performance of COMBO-17. We show that the
selection of {\it quasars}, more generally type-1 AGN, is nearly complete 
and minimally contaminated at $z=[0.5,5]$ for luminosities above $M_B =
-21.7$. Their photometric redshifts are accurate to roughly 5000~km/sec. 
Galaxy redshifts are accurate to 1\% in $\delta z/(1+z)$ at $R<21$. They
degrade in quality for progressively fainter galaxies, reaching accuracies
of 2\% for galaxies with $R \sim 22$ and of 10\% for galaxies with $R>24$.
The selection of stars is complete to $R \sim 23$, and deeper for M stars.
We also present an updated discussion of our classification technique with
maps of survey completeness, and discuss possible failures of the 
statistical classification in the faint regime at $R\ga 24$.
}
\titlerunning{A catalogue of the CDFS from COMBO-17}
\authorrunning{Wolf et al.}
\maketitle

\section{Introduction}

The Chandra Deep Field South (CDFS) is one of the most well-studied patches 
of sky. It is the target of enormous observational efforts across a wide 
range of photon energies. The variety of imaging and spectroscopic data sets 
shall improve our understanding of fundamental processes in galaxy evolution.

A large amount of public data are contributed by the {\it Great Observatories 
Origins Deep Survey} (GOODS). This survey obtains deep images of the 
field using all of NASA's great space-based facilities: the Chandra X-ray 
observatory \citep[CXO,][]{GiaCDFS}, the {\it Advanced Camera for Surveys} 
(ACS) onboard the {\it Hubble Space Telescope} \citep[HST,][]{Giav04}, and
the new infrared space telescope Spitzer. Further space-based observations 
include the {\it Ultra Deep Field} (UDF) project targetting a small part of 
the field with a single ACS pointing, deep observations with ESA's X-ray 
observatory XMM-Newton (PI Bergeron), and the wider-area ACS imaging by the 
GEMS team \citep{Rix04}. 

In this paper, we publish data and results from ground-based observations
of the CDFS. Our project, COMBO-17, has targetted the CDFS among 
four other fields. They are all observed with the {\it Wide Field Imager} 
\citep[WFI,][]{WFI1,WFI2} at the MPG/ESO 2.2m-telescope on La Silla, Chile. 
This camera covers an area of more than $0.5\degr \times 
0.5\degr$, which is larger than the field initially observed from space by 
GOODS. The footprint of this larger WFI-based image is occasionally called 
{\it Extended CDFS} or E-CDFS, but we just call it CDFS here. The purpose 
of the later GEMS images was to cover this larger area with HST resolution. 

COMBO-17 was mainly carried out to study the evolution of galaxies and 
their associated dark matter haloes at $z\la 1$ as well as the evolution
of quasars at $1\la z\la 5$. In order to obtain large samples of objects,
four fields with a total area of $\sim 1~\sq\degr$ were observed with a
17-band filter set covering the range of $\lambda_{\mathrm obs} \approx 
350 \ldots 930$~nm. This provides {\it very-low-resolution spectra} which 
allow a reliable classification into stars, galaxies and quasars as well 
as accurate photometric redshifts.

This paper publishes the full COMBO-17 catalogue \footnote{Catalogue and 
images are available at CDS and at the COMBO-17 website, 
http://www.mpia.de/COMBO/combo\_index.html}
  on the CDFS with astrometry and 17-filter photometry \footnote{In this 
paper, magnitudes are always used with reference to Vega as a zero point.}
  of 63,501 objects found on an area of $31\farcm5 \times 30\arcmin$. We also 
include classification, photometric redshifts and restframe luminosities 
 \footnote{Throughout the paper we use $H_0 = h~\times$ 100~km/(s~Mpc) in 
combination with $(\Omega_m,\Omega_\Lambda)=(0.3,0.7)$ and $h=1$ for
luminosity distances and restframe absolute magnitudes.}
  whereever the data permit their derivation. We believe, the 
classification is mostly reliable at $R\la 24$, where the sample contains
$\sim 100$ QSOs, $\sim 1000$ stars and $\sim 10000$ galaxies. Wolf et al. 
(2001c) published an earlier version of the catalogue 
containing only astrometry and BVR photometry. The version published here
contains the same set of objects with identical astrometry. However, after 
the photometry has been processed with our final procedures, we include all 
17 passbands, classifications and redshifts. 

Our catalogue could be used directly to analyse aspects of galaxy evolution, 
and some results involving more COMBO-17 fields have already been published: 
Wolf et al. (2003a) studied the evolution of the galaxy luminosity function by 
spectral type from redshift 1.2 to 0.2. Bell et al. (2004) have focussed in 
particular on understanding the red sequence evolution over this redshift. 
Accurate photometric redshifts of QSOs allowed us to observe the evolution of 
faint AGN from redshift 5 to 1 (Wolf et al. 2003b) and calculate luminosity 
functions from the largest faint and unabsorbed AGN sample to date. 

Another obvious application is the selection of sub-samples for detailed 
observations, e.g. high-resolution spectroscopy, while relying on the
knowledge of redshift and spectral type of targets. A first example drawn
from this catalogue is the measurement of velocity dispersions for $z\sim 1$ 
red sequence galaxies in the CDFS by van der Wel et al. (2004).

A number of weak lensing studies took particular advantage of the accurate 
photometric redshifts provided by COMBO-17: Kleinheinrich et al. (2004) have
used galaxy-galaxy lensing to study dark matter haloes of galaxies and their
dependence on observed galaxy properties. Gray et al. (2004) have discussed 
the correlation of galaxy properties with the underlying dark matter density 
field, based on a weak lensing mass map obtained by Gray et al. (2002). Brown 
et al. (2003) derived the shear power sepctrum and constrained cosmological 
parameters from weak lensing and redshift distributions in COMBO-17. Heymans 
et al. (2004) have later removed intrinsic alignment signals based on our 
photometric redshifts. Bacon et al. (2004) have constrained the growth of dark 
matter density fluctuations with decreasing redshift. Taylor et al. (2004) 
have demonstrated the benefit of accurate redshifts through discovering a 
background galaxy cluster in projection behind a known cluster and estimating 
its mass purely from 3-D lensing analysis. Of course, the newly discovered 
cluster could also be confirmed independently from the redshift catalogue 
itself.

A second purpose of this paper is to serve as a reference for the methodology 
of the classification and redshift estimation in COMBO-17. It is an update to
the earlier and more detailed paper by Wolf, Meisenheimer \& R\"oser (2001),
hereafter WMR. In conjunction with WMR this paper provides a full description
of the technique. We describe the performance of the classification and 
redshift estimation in the COMBO-17 data set as far as we can assess it at
this time. We assume that our catalogue will 
only be useful if we provide estimates of completeness, contamination and 
accuracy of redshifts in the star, galaxy and quasar sample. We believe that 
our redshifts for galaxies are accurate within $\sigma_z/(1+z) < 0.01$ at 
the bright end ($R<20$) where we also expect small outlier rates around 1\%. 
Redshift errors increase towards fainter levels and exceed $\sim$0.05 at 
$R>23.5$. Without NIR data faint galaxies at $z>1$ pose a tough challenge 
for our approach. We believe that the accuracy of our QSO redshifts is 
$\sigma_z/(1+z)\approx 0.015$ at all magnitudes where QSOs can be identified.

In the future, we might be able to test the quality of the photometric 
redshifts more thoroughly. The ESO team of GOODS plans a large VIMOS 
programme to obtain low-resolution spectra of more than 5000 galaxies in 
the CDFS. Such a valuable resource would allow the most systematic test of 
photometric redshifts from deep images to date. A large number of redshifts 
have already been obtained by the VIRMOS VLT Deep Survey 
\citep[VVDS,][]{LeF03} team in VIMOS GTO time.

In this paper we briefly describe the COMBO-17 observations (Sect.~2) and data 
reduction (Sect.~3), followed by an update of our classification procedure 
and template choice over WMR (Sect.~4). In Sect.~5, we present the data 
structure of the catalogue for the CDFS, and Sect.~6 gives an overview of
the object samples. In Sect.~7 we discuss completeness and redshift 
errors which we estimate from simulations of the survey. Finally, Sect.~8
discusses redshift errors in detail given the comparison with almost 1000 
spectroscopic IDs of stars, galaxies and QSOs. Whenever we present
numbers in this paper, we refer specifically to the CDFS dataset as it
is published here, but whenever we talk about techniques in general, 
they are applied to all fields observed in COMBO-17.

\section{Observations}

\begin{figure*}
\centering
\includegraphics[clip,angle=270,width=15cm]{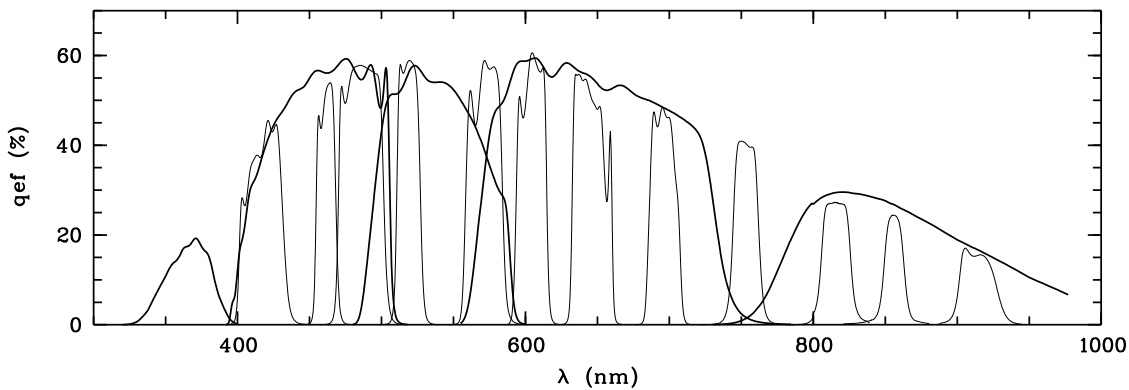}
\caption{COMBO-17 filter set: Total system efficiencies in the COMBO-17 bands. 
They include two telescope mirrors, the WFI instrument, CCD detector and an 
average La Silla atmosphere. Photometric calibrations of such datasets are 
best achieved with spectrophotometric standards inside the target field. }
\label{qeff}
\end{figure*}

\begin{table*}
\caption{COMBO-17 imaging data on the CDFS: For all filters we list the total 
exposure time, the PSF on the co-added frames, the 10$\sigma$ (Vega) magnitude 
limits for point sources and the observing runs (see Tab.~\ref{obsruns})
in which the exposure was collected. For flux and magnitude conversions
we list the AB magnitudes and photon fluxes of Vega in all our filters. 
The $R$-band observations were taken in the best seeing conditions. 
\label{filterset}}
\begin{tabular}{ccrccl|cc}
\hline \noalign{\smallskip} \hline \noalign{\smallskip} 
  \multicolumn{2}{c}{$\lambda_\mathrm{\mathrm{cen}}$/fwhm}  & 
  $t_\mathrm{\mathrm{exp}}$  &  seeing  & 
  $m_\mathrm{\mathrm{lim},10\sigma}$  &  run code  &
  mag of Vega  &  $F_\mathrm{phot}$ of Vega \\
  \multicolumn{2}{c}{(nm)}  &  (sec)  &   &  (Vega mags)  & 
     & (AB mags)   &  $(10^8~\mathrm{photons/m^2/nm/s})$  \\ 
\noalign{\smallskip} \hline \noalign{\smallskip} 
365/36  &$U$ & 21600 & 1\farcs00  & 23.9 & F, G  & $+0.77$ & 0.737\\ 
458/97  &$B$ & 11240 & 1\farcs10  & 25.1 & D, F  & $-0.13$ & 1.371\\ 
538/89  &$V$ &  8400 & 1\farcs20  & 24.5 & D     & $-0.02$ & 1.055\\
648/160 &$R$ & 35700 & 0\farcs75  & 25.4 & D, E, F, G&$+0.19$&0.725\\ 
857/147 &$I$ &  9800 & 1\farcs20  & 22.5 & D     & $+0.49$ & 0.412\\
\noalign{\smallskip} 
418/27 &   &  7900 & 1\farcs20  & 24.0 & E     & $-0.19$ & 1.571\\ 
462/13 &   & 10000 & 0\farcs95  & 24.1 & E     & $-0.18$ & 1.412\\ 
486/31 &   &  5000 & 1\farcs30  & 23.6 & D     & $-0.06$ & 1.207\\ 
519/16 &   &  6000 & 1\farcs15  & 23.6 & E     & $-0.06$ & 1.125\\ 
572/25 &   &  4750 & 0\farcs95  & 23.5 & D, E  & $+0.04$ & 0.932\\ 
605/21 &   &  6000 & 1\farcs00  & 23.5 & E     & $+0.10$ & 0.832\\ 
645/30 &   &  4000 & 1\farcs15  & 22.4 & D     & $+0.22$ & 0.703\\
696/21 &   &  6000 & 0\farcs90  & 23.2 & E     & $+0.27$ & 0.621\\ 
753/18 &   &  7000 & 1\farcs00  & 22.5 & E     & $+0.36$ & 0.525\\ 
816/21 &   & 20000 & 1\farcs00  & 22.9 & E, G  & $+0.45$ & 0.442\\ 
857/15 &   & 16560 & 0\farcs90  & 22.0 & D     & $+0.56$ & 0.386\\ 
914/26 &   & 15200 & 1\farcs10  & 21.9 & D, E  & $+0.50$ & 0.380\\
\noalign{\smallskip} \hline
\end{tabular}
\end{table*}

The COMBO-17 survey has produced multi-colour data in 17 optical filters on 
1~$\sq\degr$ of sky at high galactic latitudes. The survey includes the 
Chandra Deep Field South, centered on the coordinates $\alpha_{2000} = 
03^{\mathrm{h}} 32^{\mathrm{m}} 25^{\mathrm{s}}$ and $\delta_{2000} = 
-27\degr 48\arcmin 50\arcsec$ (see Wolf et al. 2003 for details regarding
the other fields). The filter set (Fig.~\ref{qeff} and Tab.~\ref{filterset}) 
contains five broad-band filters (UBVRI) and 12 medium-band filters covering 
wavelengths from 400 to 930~nm. In this paper, we focus on our observations 
of the CDFS.

All observations presented were obtained with the Wide Field Imager at the 
MPG/ESO 2.2m-telescope on La Silla, Chile. The WFI provides a field of view 
of $34\arcmin \times 33\arcmin$ on a CCD mosaic consisting of eight 2k 
$\times$ 4k CCDs with at a scale of $0\farcs238$
per pixel. The observations on the CDFS were spread out over four independent
observing runs between October 1999 and February 2001. They encompass a total 
exposure time of $\sim$195~ksec of which $\sim$35~ksec were taken in the 
R-band during the best seeing conditions. 

We needed several observing runs to collect the full 17-filter data set.
Hence, for some long-term variable stars and QSOs, the observed
17-passband spectral energy distribution (SED) might be skewed. We
attempt to ameliorate the color-independent part of the long-term
variability by taking at least R-band data for every observing run, 
with which we can normalise the SED. We can not correct for variability in 
colours or for short-term variability on time-scales within an observing 
run. In addition to long exposures for efficient light gathering, we 
included short exposures for the photometry of bright objects, in particular
to avoid saturation of our brighter standard stars in broad filters.

The long exposures followed a dither pattern with ten telescope pointings 
spread by $\Delta\alpha$, $\Delta\delta <\pm 72\arcsec$. The pattern allows 
us to cover the gaps in the CCD mosaic, but also minimises field rotation. 
Owing to the gaps in the CCD mosaic the total exposure time varies from 
pixel to pixel. However, the dither pattern makes each position on the sky 
fall onto a CCD in at least eight of ten exposures, while 97\% of all sky 
pixels are recorded in ten out of ten farmes.

Twilight flatfields were obtained with offsets of $10\arcsec$ between
consecutive exposures. Exposure times ranged between 0.5 and 100 seconds
per frame. Note that the WFI shutter design allows exposures as short 
as 0.1 seconds without causing significant spatial variations in the 
illumination across the CCD mosaic \citep{Wackermann}.

\begin{table}
\caption{COMBO-17 observing runs with CDFS imaging. 
\label{obsruns} }
\begin{tabular}{ll}
\hline \noalign{\smallskip} \hline \noalign{\smallskip} 
run code    &  Dates  \\ 
\noalign{\smallskip} \hline \noalign{\smallskip}
D           &  07.10.-22.10.1999  \\
E           &  28.01.-11.02.2000  \\
F           &  21.09.-30.09.2000  \\
G           &  19.01.-20.01.2001  \\
\noalign{\smallskip} \hline
\end{tabular}
\end{table}

We have established our own set of tertiary standard stars based on 
\emph{spectrophotometric} observations, mainly in order to achieve a 
homogeneous photometric calibration for all 17 WFI filter bands. Two stars 
of spectral types F/G and magnitudes $B_J \simeq 16$ were selected in 
each COMBO-17 field, drawn from the Hamburg/ESO Survey database of digital 
objective prism spectra \citep{Wis00}. The spectrophotometric 
observations for the CDFS were conducted at La Silla on Oct 25, 1999,
using the Danish 1.54\,m telescope equipped with DFOSC. A wide ($5\arcsec$)
slit was used for the COMBO-17 standards as well as for the external
calibrator, in this case the HST standard HD~49798 \citep{bohlin92}. 
Two exposures of 45~min were taken of each star, one with the blue-sensitive 
grism 4 covering the range $\lambda=340$--740~nm, and one with the 
red-sensitive grism 5 covering $\lambda>520$~nm.

\section{Data reduction}

\subsection{Image reduction and object search}

All procedures used for the data reduction are based on the MIDAS package. 
A WFI image processing pipeline was developed by Wolf et al. (2001) and makes 
intensive use of programmes developed by K. Meisenheimer, H.-J. R\"oser and 
H. Hippelein  for the Calar Alto Deep Imaging Survey (CADIS). The pipeline 
performes basic image reduction and standard operations of bias subtraction, 
CCD non-linearity correction, flatfielding, masking of hot pixels and bad 
columns, subtraction of fringe patterns, cosmic ray rejection and subsequent 
stacking into a deep co-added frame that is common to all dither pointings. 

Our deepest co-added frame is the $R$-band image obtained in run D which has
a uniform, sharp PSF with $0\farcs75$ FWHM. It provides the most sensitive 
surface brightness limits and the highest signal-to-noise ratio for object 
detection and astrometry among all data available in the survey. Only 
L~stars and quasars with $z>5$ have higher S/N in redder bands, a fact 
which we ignore at this stage. The WFI field of view and the dither pattern 
lead to a common area of $31\farcm47\times 30\farcm11$ in the $R$-band image. 

We used the SExtractor software \citep{BA96} with default setups 
in the parameter file, except for choosing a minimum of 12 significant 
pixels required for the detection of an object. We first search rather deep 
and then clean the list of sextracted objects from those having a S/N ratio
below 3, which corresponds to $>0\fm3123$ error in the total magnitude
MAG--BEST. As a result we obtained a catalogue of 63,501 objects in the 
CDFS, with positions, morphology, total $R$-band magnitude and its error, 
reaching a 5$\sigma$ point source limit of $R\approx 26$. In this paper, 
magnitudes are always cited with reference to Vega as a zero point. The
astrometric accuracy appears to be better than $0\farcs15$ (see also the
comparison with the astrometry from $H$-band imaging by Moy et al., 2003).

\subsection{Spectral energy distributions}

The spectral shapes of the objects in the $R$-band selected catalogue were 
measured with a different approach. Photometry in all 17 passbands was done
by projecting the object coordinates into the frames of reference of each 
single exposure and measuring the object fluxes at the given locations. In 
order to optimize the signal-to-noise ratio, we measure the spectral shape
in the high surface brightness regions of the objects and ignore potential 
low surface brightness features at large distance from the center. However,
this implies that for large galaxies at low redshifts $z<0.2$ we measure 
the SED of the central region and ignore colour gradients. 
 
\begin{figure}
\centering
\includegraphics[clip,angle=270,width=0.95\hsize]{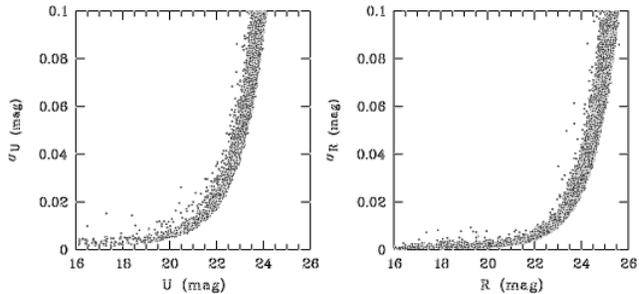}
\caption{Errors versus magnitudes for all objects with errors below
$0\fm1$ in the WFI $U$ and $R$ filter (combined photometry from all 
epochs). 
\label{errors}}
\end{figure}

Since seeing variations among the different bands would introduce artificial 
colour offsets and changing observing conditions are typical for ground-based 
observations, we need a non-standard photometry approach to measure spectral
shapes accurately. In fact, we need to measure the same central fraction of 
an object in every band as it would appear in equal seeing. To this end, we
use the seeing-adaptive, weighted aperture photometry in the MPIAPHOT package 
\citep{RM91,Mei03}.  

MPIAPHOT measures the central surface brightness of objects after convolving 
their appearance outside the atmosphere to an effective PSF of $1\farcs5$ 
diameter. In detail, the procedure measures the observed stellar PSF on each 
individual frame and chooses the necessary Gaussian smoothing for reaching a
common effective PSF of $1\farcs5$ uniformly on all frames in all bands. For
most objects this measurement is similar to a flux measurement in an aperture 
of $1\farcs5$ diameter in $1\farcs5$ seeing.

The fluxes from individual frames are averaged into a final flux for each 
object with the flux error being derived from the scatter among the frames. 
This way, the error does not only take photon noise into account, but also 
systematic effects, such as suboptimal flatfielding and uncorrected CCD 
artifacts, which can only be included in the error budget because the object 
location on the CCD varies with dithering. Also, in this final averaging step 
we prevent chance coincidences of count rates from pretending unreasonably 
low errors by using the errors derived from background and photon noise as a 
lower limit (see Meisenheimer et al., in prep, for a full discussion of the 
photometric analysis). Furthermore, we do not want transient bad pixels or 
columns to affect object fluxes, when they are not contained in our bad pixel 
mask. We eliminate the most outlying single measurement if that reduces the 
scatter significantly.

\subsection{Photometric calibration}

The photometric calibration is based on our own spectrophotometric standard 
stars within the field. We have calibrated these with respect to external 
standard stars in photometric nights. As a result, we are independent from 
photometric conditions for all imaging.

The spectra were reduced by standard procedures and have a final
signal-to-noise ratio of $> 30$ per pixel except very near to the low-
and high-wavelength cutoffs. The agreement between spectra in the
substantial overlap in wavelength between the two grisms is excellent,
confirming that contamination from second order was negligible. We
estimated the absolute spectrophotometric accuracy by comparing several
spectra of the external calibrator HD~49798 obtained during the entire
observing run. The little variation of the overall flux levels indicates 
that the magnitude calibration is better than 10\%. We further estimate
that the relative calibration between various wavebands is better than 5\%. 

The flux calibration for the whole catalogue is then achieved by convolving 
the spectra of our standard stars with the total system efficiency in our 
filters. We then know the physical photon flux we have to assign to them, 
and establish the flux scale for all objects. The validity of the relative
calibration, i.e. the shape of the SED, was finally confirmed by comparing 
the observed stellar locus with its synthetic colours.

Unfortunately, we ended up having only one calibration star in the CDFS,
which is furthermore almost located at the edge of the field. The data of
the other, more central star in the CDFS turned out to be faulty after the
spectroscopic observations were finished. So, the established flux scale
relies on object 60873 at $\alpha_{2000} = 03^{\mathrm{h}} 32^{\mathrm{m}} 
02\fs360$ and $\delta_{2000} = -27\degr 34\arcmin 22\farcs1$. This star 
of $R\approx 15.5$ matches very well the template of an F5V star from 
Pickles (1998).

\subsection{Variability correction}

Since the multi-colour observations were collected over two years and some 
objects show variability, it was necessary to correct for the latter when 
constructing the 17-filter spectra for classification. Otherwise, the 
non-simultaneous SED could mislead the classification about the nature of 
the object and its photometric redshift. Indeed, the main variable objects 
are quasars and Seyfert galaxies, but also a few stars and Supernovae 
superimposed on galaxies. We note, that tests ignoring the variability of 
quasars have dramatically increased their photometric redshift errors.

Over the entire observing period of COMBO-17 we collected deep $R$-band data 
so we can measure variability at least in the $R$-band. When constructing the
SEDs of variable objects, we relate the measurements of every observed band
to the $R$-band magnitude obtained in the same observing run. As a result, 
the SEDs are not distorted by long-term magnitude changes. This variability
correction does not take changes in spectral shape into account, and we also
can not correct any short-term variability on the time scale within an
observing run as we do not have continuous $R$-band monitoring available.

In search for variability we calculated magnitude differences between any two 
flux measurements $F_1\pm\sigma_1$ and $F_2\pm\sigma_2$ in the same filter 
\begin{equation}
   \Delta m = 2.5 \log{F_1 / F_2}  ~ , 
\end{equation}

and the associated error 
\begin{equation}
   \sigma_\Delta = \frac{1.085}{\rm{max}(F_1,F_2)}
                   \sqrt{\sigma_1^2 + \sigma_2^2}  ~ .
\end{equation}

This comparison was possible for the $R$-band which was observed in all four
observing runs, but also in five further bands which were observed twice 
(see Tab.~\ref{filterset}). 
We take a conservative approach to flagging objects as variable, where we
only want to identify genuinely variable objects and avoid contamination
by noise spikes. We use a rather high threshold for considering a change
significant, because among 60,000 objects we would still expect $\sim150$ 
spurious outliers above, e.g., a $3\sigma$ threshold. Also, we are not 
interested in finding variables among very faint objects, where we can not
obtain any redshift estimates, so we used a signal-to-noise cut on the
brightness as well. In the end, we only flagged objects as variable when 
all three of the following criteria were met:
\begin{itemize}
\item at least one of the measurements was determined to better than 
 10$\sigma$, i.e. $\sigma_1<0.1$ or $\sigma_2<0.1$
\item the difference was at least $0\fm15$, i.e. $\Delta m > 0.15$
\item the difference had at least 6$\sigma$ significance, 
  $\Delta m/\sigma_\Delta > 6$ 
\end{itemize}
 
We applied the variability correction of the SED only for flagged objects.
The derived errors of the corrected SED include also the recalibration 
uncertainties from the $R$-band observations in the various runs. If we 
applied corrections to all objects, we would increase the total errors in 
the SED determination of non-variable objects due to the intrinsic error of 
the correction itself. For objects not fulfilling the variability criteria, 
we averaged all available measurements in a given filter as measured.

\subsection{Flags}

The final catalogue contains quality flags for all objects in an integer 
column, holding the original SExtractor flags in bit 0 to 7, corresponding 
to values from 0 to 128, as well as some internal quality control flags of 
our photometry in bits 9 to 11 (values from 512 to 2048). The meaning of 
SExtractor flags can be found in SExtractor manuals. We generally recommend 
users to ignore objects with flag values $\ge 8$ (i.e. any bit higher than 
bit 2 is set), for any statistical analysis of the object population. 
Their photometry might be affected by various problems, e.g. saturation.

If an object of interest identified in another data source shows bad 
flags here, it may still have accurate COMBO-17 photometry. Bit 9 
indicates only a potential problem (check images for suspicious bright 
neighbours or reflections from the optics). Higher bits reflect fatal 
errors: Bit 10 is set when uncorrected hot pixels have severely affected 
the photometry, and bit 11 was set for some objects we had interactively 
identified to have erroneous photometry.

\section{Classification and redshift estimation}

The interpretation of our object SEDs or {\it very-low resolution spectra}
involves a classification and redshift estimation. It is performed by an
automated procedure, as it is the case for any photo-z's by other authors
and for redshift determination in large and modern spectroscopic surveys. 
A number of authors prefer to reserve the term {\it photo-z} to photometry 
data of low spectral resolution as it is obtained in broad-band surveys. 
Hence, it would be inappropriate for medium-band redshifts in COMBO-17.
Koo (1999), e.g., presented a brilliant review on the subject of photometric
redshifts, and proposes to limit the term to imaging data from filters
with $\lambda/\Delta\lambda \la 20$.

But the term {\it spectroscopic redshifts} would also be unsuitable here.
Present-day techniques for constraining redshifts in either photometric or 
spectroscopic surveys, including COMBO-17, are mathematically all similar.
Whether they are applied to 5 passbands, 17 passbands, 500 spectral channels,
or even to 2 passbands, makes only a difference in the discriminative power
of the data. In the past, we have started to call our medium-band approach 
by the name {\it fuzzy spectroscopy}. 

The underlying mathematics of statistical classification procedure have 
been discussed in WMR. For all details of the technique not explained here, 
we refer the reader to that source. That work has also modelled surveys 
with filter sets of different width while keeping the total survey exposure 
time constant. The results demonstrated that medium-band surveys with less 
exposure per filter are obviously less deep in terms of object detection 
than broad-band surveys. But surprisingly they reach to an equal depth in 
terms of the subsample of objects which are successfully classified and 
have useful redshift estimates. This means that medium-band surveys produce 
a similar amount of information as broad-band surveys. At brighter 
magnitudes, they have the advantage of delivering more information due 
to the larger number of independent bands, resulting in higher redshift 
accuracy: a sufficient motivation for the COMBO-17 survey.

\subsection{Historical remarks}

The idea of photometric redshifts for galaxies dates back to the 1940's, 
when Messrs Baade and Hubble initiated a programme to extend tests of 
cosmological models beyond spectrographic limits and use the potential 
of the 200'' telescope to its photographic limit instead \cite{SW48}. 

Using photoelectric detectors in nine bands from 350 to 1050 nm and just
a single template for elliptical galaxies, Baum (1962) measured redshifts 
of cluster galaxies out to $z\sim 0.5$ with remarkable accuracy, on the 
order of $\sigma_z < 0.02$. His IAU presentation (see discussion in Baum 
1962) was met with a mix of encouragement and scepticism. Much that has 
been said about photo-z's in the 1950's and 60's, still applies today:

\begin{itemize}
\item
   Photo-z's are controversial: Typically, as redshift estimates they
   are useful for large samples of objects, but for an individual 
   object they are no proof of its nature.
\item
   Photo-z's support spectroscopic observations, e.g. when only one line
   is identified or features are weak and more information is needed.
\item
   Spectroscopic checks are usually appreciated to validate a photo-z
   sample to ascertain that it is free from systematic errors. 
\item
   Systematic errors arising from calibration issues and from mismatch
   between real spectra and assumed templates are the main limitations
   to photo-z performance, while aiming for great photon statistics is
   usually irrelevant.
\end{itemize}

The redshift accuracies published in the literature have not improved 
over the last 40 years, but by now the technique is no longer restricted 
to elliptical galaxies. Therefore, when we say, that {\it photometric 
redshifts have developed into a mature tool of astronomical information 
extraction}, we mostly mean that they can be more universally applied.
Today, photo'zs allow simple spectral classification and even provide 
redshifts for quasars \citep{Wolf99,Bud01}.

Photo-z's are more reliable when more spectral bands are available and 
when they cover a wider wavelength base. Increasing the wavelength
resolution of the filter set increases the redshift accuracy. For a
brilliant review on the subject of photometric redshifts, see Koo (1999),

Usually, photometric redshifts are obtained for large object catalogues 
to provide galaxy or AGN samples for follow-up analysis. Such samples
typically spread over a wide range of object magnitudes across which 
the redshift errors, completeness and contamination change. A simple 
picture splits the magnitude range into three quality domains, which 
we specify here for the COMBO-17 survey:

\begin{enumerate}
\item 
  the {\it quality saturation domain} at $R<22$, where systematic errors
  from calibration and especially from poor match between templates and
  true SEDs of celestial objects dominates the performance while the 
  collected photon statistics is irrelevant
\item
  the {\it quality transition domain} at $R=22\ldots 24.5$, where photon
  statistics take over degrading the performance with decreasing signal
\item
  the {\it quality breakdown domain} at $R>24.5$, where the classification
  is undecidable using the collected photons and a-priori knowledge of what
  we should expect dominates the interpretation of the objects completely.
\end{enumerate}

\begin{figure*}
\centering
\hbox{
\includegraphics[clip,angle=270,width=0.5\hsize]{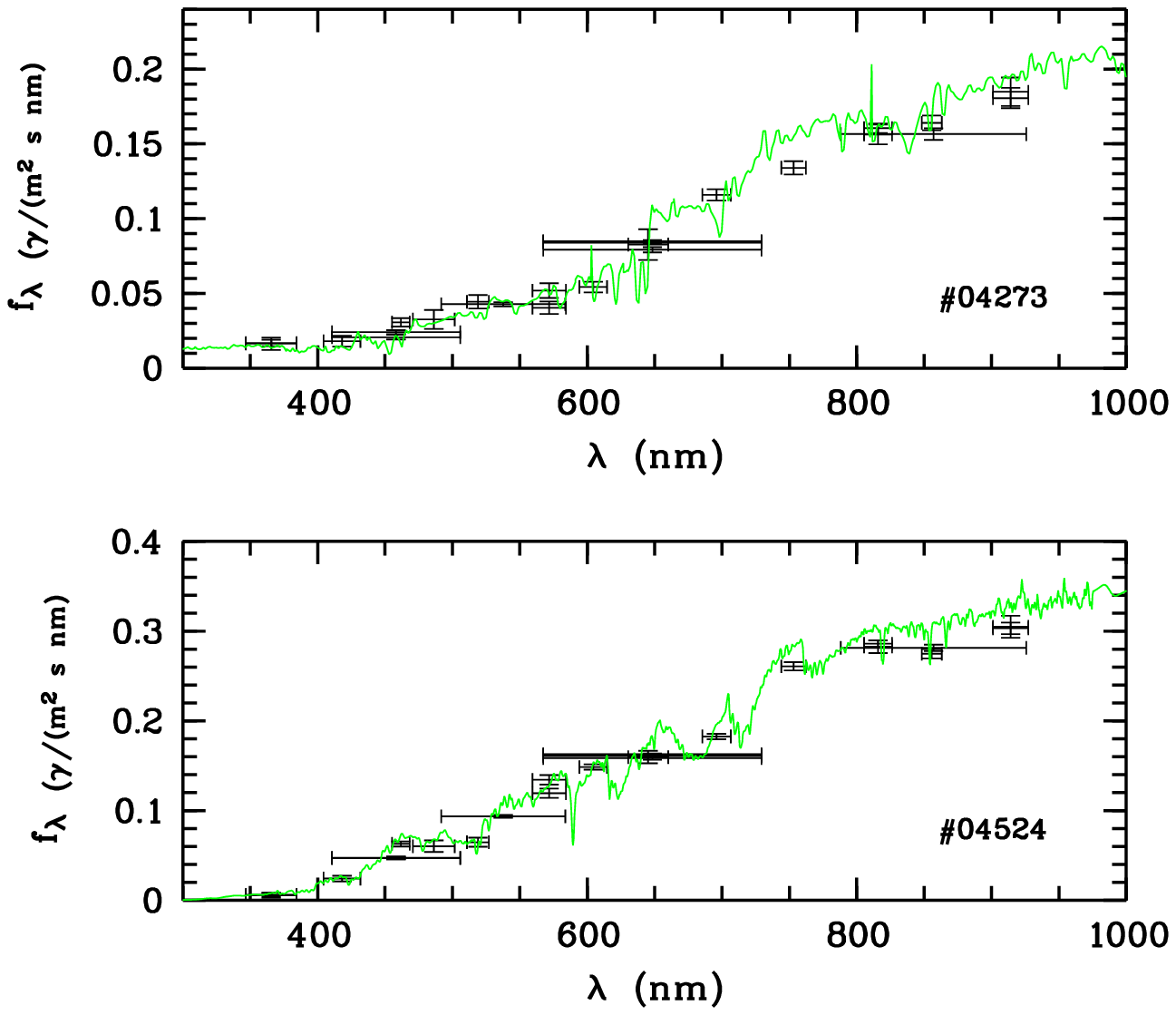}
\includegraphics[clip,angle=270,width=0.5\hsize]{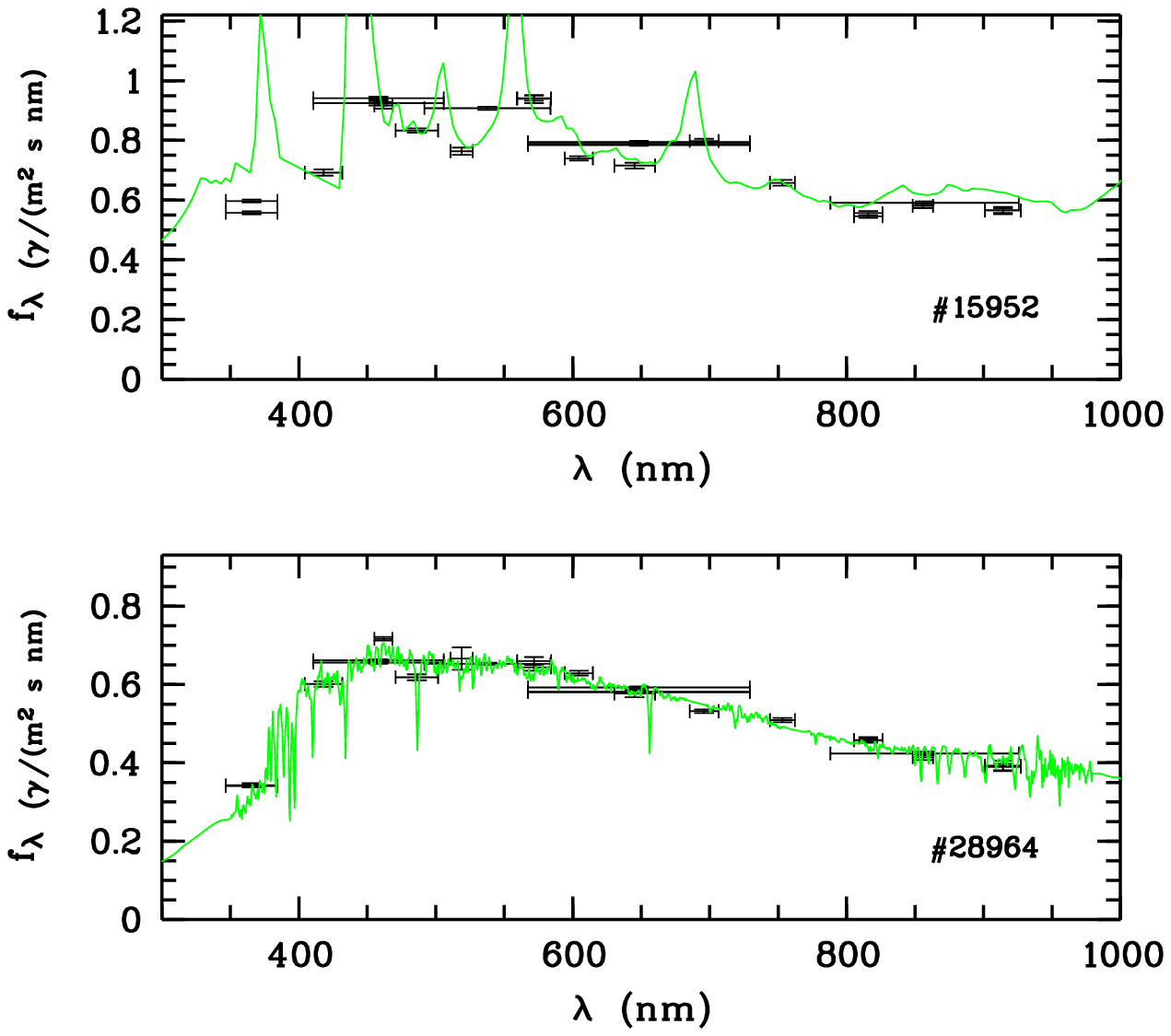}}
\caption{{\it Medium-band power:} Pair comparison between objects with 
similar $BVRI$ colours, where different medium-band SEDs break the 
degeneracies in broad-band colour diagrams. Multiple error bars mean 
multiple observations at different epochs.
{\it Left panels:} A $z=0.67$ galaxy with modest star formation (top) 
and an M1 star (bottom). 
{\it Right panels:} A $z=2.6$ quasar (top) and an F2 star (bottom).
\label{compSEDs}}
\end{figure*}

\subsection{Basic technique}

One basic technique of {\it photo-zeeing} is to use a template database 
as a model fit to the colour data of an observed object. The templates 
are arranged into a structured grid with model parameters such as redshift, 
extinction or SED type. As in any model fit, the model needs to be a good 
description of the data to yield sensible results, otherwise the 
statistical test will give a misleading result. So, implicitly we assume 
that we know of all possible spectra in the universe, and calculate the 
probability for each of them to produce the colours of an observed object.

The validity of any model fit also depends on the use of correct errors.
We emphasize that our method for deriving robust error estimates for each
object (see above) is essential to get an accurate statistical confidence 
for our results, e.g. for the class probabilities and for the estimation 
of redshift errors of each object itself. We also have to ascertain that 
there is no mismatch between observed data and template models due to 
calibration errors. The relative calibration between different passbands 
is uncertain at a 3\% level, and templates may mismatch real SED data at 
a similar level.

Our template fits do not operate on a linear flux scale but in a colour 
space of colour indices as detailed in WMR. In order to take calibration 
tolerances and template mismatch into account we assume a minimum error 
for each colour index. To this end, we add an error floor of $0\fm05$ 
quadratically to every index error. As in WMR, we calculate explicit 
colour libraries for a grid in redshift from the spectral libraries 
before starting any classification code on object catalogues, which 
saves vast amounts of computing time. 

We like to comment on possible other photo-z techniques here: the most
relevant alternative to template fitting is based on fitting the empirical
colour distributions of galaxies or quasars. This is a valuable approach 
whenever good model descriptions are unavailable, but there is a large 
training set of objects with known types and redshifts. It involves 
either explicit fitting of low-order polynomials \cite{Con95} or implicit 
fitting as in Artificial Neural Nets \citep[ANNs, e.g.][]{FLS03}.
In our case of a rather deep survey reaching to $z\ga1$ for galaxies, the
classical method of template fitting is currently superior. Once a large 
number of spectroscopic identifications across all the redshifts become
available, ANN approaches will be feasible for COMBO-17 as well.

\subsection{Spectral templates}

In this subsection we give an update on our current choice of templates
over the detailed discussion in WMR.

\subsubsection{Star templates}

We still take our star templates from the spectral atlas of Pickles (1998),
but we restrict it to the spectral types FGKM, because we do not expect 
any main-sequence OBA stars in this field. Instead we predict from Galactic 
Halo models that we should find about one Blue Horizontal Branch (BHB) star,
several white dwarfs and possibly blue (sdB) subdwarfs.
For the latter, we introduced a new class for blue high-gravity stars 
using synthetic templates by Koester (priv. comm.), effectively matching
colours of white dwarfs and sdB subdwarfs, but also matching BHB stars 
in the low-$g$ domain. From Koester's grid of atmosphere models we select
the DA dwarf models in the temperature range from 6,000~K to 40,000~K
and the surface gravity range of $\log g = 6.0\ldots 9.0$.

\begin{figure}
\centering
\includegraphics[clip,angle=270,width=\hsize]{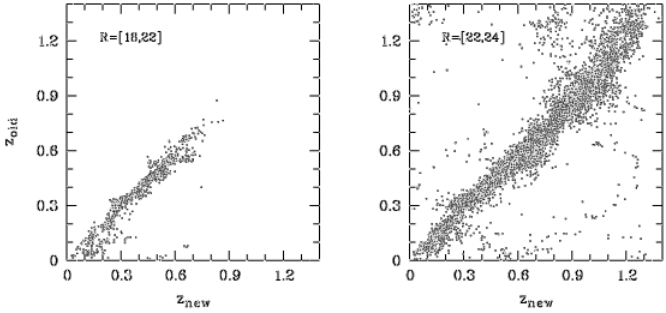}
\caption{Galaxy redshifts old and new: The change-over in the template
set for galaxies has changed the redshifts of most galaxies relatively
little. But it has improved the overall redshift accuracy and also the
completeness in redshift determination.
\label{zoldz}}
\end{figure}

\subsubsection{Galaxy templates}

The galaxy template library used in our team has been completely rebuilt 
and contains now a grid of synthetic spectra based on the PEGASE code
\cite{FRV97} for population synthesis models. In the past, we used the 
observed templates by Kinney et al. (1996), which where still used for the 
work on the evolution of the galaxy luminosity function (Wolf et al. 2003a). 

We found that the new templates deliver a better redshift accuracy than the
old ones. The changes in redshift are relatively small as demonstrated in 
Fig.~\ref{zoldz} and within the errors of the old redshift estimates. But 
the residual errors have been reduced in the new optimised redshifts and
allow unprecedented photo-z accuracy. The resulting luminosity functions 
of galaxies are unchanged within the errors published in Wolf et al. 
(2003a). The luminosity functions of red-sequence galaxies published by
Bell et al. (2004) are already based on the new redshifts.

The new grid includes the full range of the Kinney templates, and actually 
covers a wider range of spectral types, extending to slightly bluer and 
redder spectra. Further advantages of the synthetic spectra are the more 
regular grid in spectral types, the more physical parametrization of the 
types and the possibility to extend the grid into more than just one 
dimension given by a single type parameter.

The templates span a two-dimensional grid with a range of ages calculated 
by the PEGASE code and a range of extinction levels which we applied as 
external screens to the SEDs delivered by PEGASE. The setup for PEGASE uses 
standard parameters suggested by the codes {\it SSPs} and {\it scenarios} 
with a Kroupa (1993) IMF and no extinction. The star formation history 
follows an exponential decay law with a time constant of $\tau=1$~Gyr. 
The SEDs are calculated by PEGASE for various time steps since the 
beginning of the first star formation. As templates we use 60 spectra for 
look-back times (''ages'') ranging from 50~Myr to 15~Gyr.

We tune initial model metallicities to give almost solar metallicity
over the whole range of templates. This reproduces approximately the
metallicity of the L* galaxies which dominate any magnitude-limited sample. 
It is worth noting, however, that the well-known age/metallicity degeneracy 
is actually helpful in this case: Mismatches between real galaxy and 
template galaxy metallicities can be compensated for by modest changes in 
template age, while yielding nonetheless accurate estimates of redshift 
and SED shape.

The COMBO-17 classifier also allows for dust reddening. The COMBO-17
photometry of galaxies with known redshift $z>0.6$ shows little sign of 
an absorption trough near restframe 220~nm. In contrats, a Milky Way-type 
extinction law has a strong 220nm trough, and hence we do not adopt such 
a reddening law for our templates. We experimented with the well-known 
Calzetti law, but it seemed to produce insufficient curvature between 180 
and 400~nm to match our observations. We have found a satisfactory ad-hoc 
solution using the 3-component extinction law by Pei (1992). We decided to 
use his SMC law, because it provided a reasonably good match between 
templates and observed SEDs of galaxies with previously known redshift. The 
issue of dust extinction certainly deserves a more thorough exploration, 
for which COMBO-17 could supply a wealth of photometric data. However, we 
defer this issue to a future work. Note, that the detailed choice of 
extinction law has no effect on the observed SEDs of galaxies with $z<0.6$. 
All templates are then extinguished with six different equidistant degrees 
of reddening in the interval of $E_{B-V}=[0.0,0.1,...0.5]$.

The redshift grid for the galaxy colour library covers the range from
$z=0$ to $z=1.40$ in 177 steps. These are equidistant on a $\log (1+z)$
scale with steps of 0.005 and of course limit the redshift resolution 
when reconstructing galaxy density features in redshift space. According 
to the sampling theorem, features in redshift space can be recovered if
their wavelength is at least as large as two grid steps. Thus, we have
to expect that features with wavelength $\delta_z/(1+z) < 0.01$ will be
smoothed by our redshift estimation even under perfect conditions where
systematic problems in photometric calibration or SED match are absent.
This will not significantly restrict the power of our dataset, 
because our redshift estimation is probably not consistently much more 
accurate than 0.01 across all redshifts and SEDs we are interested in.

\subsubsection{QSO templates}

The QSO template library is derived from the SDSS template spectrum 
\cite{vdB01}. This template resembles a typical emission line contour 
on top of an average QSO continuum. In order to cover the range of SEDs
expected for QSOs, we generated spectra of different continuum slope 
and emission line strength. For this purpose, we have varied the given
template in intensity and added it to a power-law continuum. We do not
need to vary the relative strength of different emission lines by much,
since the filter set usually shows one, or sometimes two, emission lines
in the medium-band filters. Finally, we have taken the Hydrogen absorption 
bluewards of the Lyman-$\alpha$ line into account by multiplying intrinsic 
SEDs by a redshift-dependent throughput function \cite{MJ90}. 

The effective spectral indices of the resulting spectra depend on redshift 
since the continua are not power laws. The observed $B-I$ colour of the QSO 
templates at $z=2$ runs from about +0.35 to +1.75 (in Vega magnitudes), 
corresponding roughly to power law indices from $\alpha=-1.66$ to $+0.4$. 
The final colour library covers QSO redshifts across $z=[0.5,5.96]$. At 
$z<0.5$, the template does not cover the full COMBO-17 filter set, so that 
the fits would be less constrained than at other redshifts. Hence, the
classifier is insensitive to QSOs at $z<0.5$. In fact, we expect very few
such objects, and chose to ignore the problem. The library grid has 20 steps 
in the spectral slope axis, nine steps in emission line strength and 155 
steps in redshift with a resolution of 0.01 in $\log (1+z)$. Hence, features 
in $z$ space smaller than $\delta_z/(1+z)=0.02$ are smoothed, but we do 
not intend to study such substructure in the small QSO sample, anyway.

\begin{figure}
\centering
\includegraphics[clip,angle=270,width=\hsize]{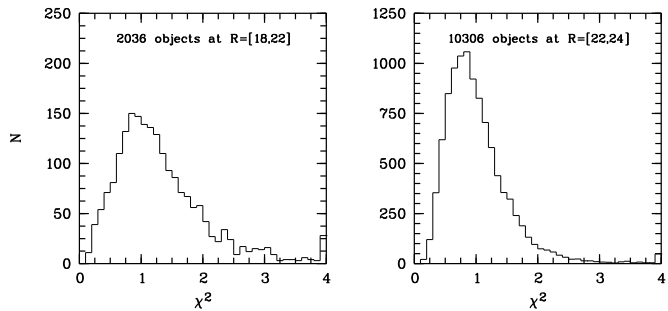}
\caption{Template fit -- histograms of reduced $\chi^2$ values: 
{\it Left panel:} At 
bright levels of $R=[18,22]$ photon noise is irrelevant. The quality of 
template fits is limited by calibration errors and template variety. The 
$\chi^2$ values scatter a bit higher than expected. {\it Right panel:}
For fainter objects, flux errors from photon noise dominate the template 
fits. Hence, they show very much the expected $\chi^2$ distribution.
\label{chi2red}}
\end{figure}

\subsection{Classification rules}

First, the probability of each template producing the observed colours of
a given object is calculated on the basis of the match quality and the
photometric errors. Then, its actual classification involves a decision 
between the basic alternatives {\it Star}, {\it WDwarf}, {\it Galaxy} and 
{\it QSO}. Therefore, we compare the integrated probabilities of each of 
the four classes, after normalising their sum to 100\% to indicate our 
assumption that no further class may exist.

Before the decision is made, we multiply each class probability with the 
a-priori probability of an object to belong to any of the classes depending
on its observed magnitude (in the I-band). Hence, we have used all the
available photometric information for the classification. The first step
of template comparison uses only colour indices or spectral shapes, but
in the second step the overall brightness (i.e. the free normalisation
parameter in template fitting) is factored into the probabilities via 
a-priori class distributions. These a-priori functions are simple linear
approximations of our number counts. At the faintest levels, where the
classification breaks down, the number counts are uncertain, and the
linear fits probably don't reflect the true distribution.

As in WMR, a decision is made in favour 
of a single class if it accounts for $>75$\% of the total probability.
At $R<23$, around 90\% of the objects are classified with a probability 
of $>99$\% focussed on one class interpretation, suggesting a 1-in-100 
risk for misclassifications among these. The other 10\% are less clear
cases, and reside in parts of the colour space where templates from more 
than one class crowd together, implying the filter set is not sufficient 
to discriminate between them unambiguously. A decision based on 75\% 
probability in favour of one class translates into a 1-in-4 chance that 
the wrong class has been assigned! We assign {\it Galaxy (Uncl!)} to
unclassified objects below the 75\% margin, reflecting that the decision 
is very uncertain. Most often, these objects are galaxies anyway, because
the linear number count fits used by us tend to overestimate the density
of stars and QSOs at faintest levels. In any case, it will be the safest 
approach to count those in the rich {\it Galaxy} class if anywhere. 
Some true stars and QSOs will be in this group, but they do not form a 
significant fraction of contaminants given how dominant galaxies are at 
faint magnitudes (see also discussion on contamination in Sect.~7.3).

There are a few conditions, under which we override this statistical 
approach with hard rules:

Faint stars at the red end of the {\it WDwarf} grid or on the blue end 
of the Pickles FGKM sequence should be classified either as {\it Star} 
or {\it WDwarf}. But their large photometric errors can result in high 
probabilities just below 50\% for both classes, and we do not want them 
to be labelled {\it Galaxy (Unclassified)}. So, if $p(Star)+p(WDwarf) > 
75$\% we still assign the more probable one of the two stellar classes 
with no further comment.

A second non-statistical rule is applied for objects with point-source 
morphology and classification as galaxies with $z<0.2$. Noisy photometry 
leaves faint galaxies at redshifts near zero difficult to differentiate 
from stars and sometimes causes their misclassification as stars. In case 
of a clearly extended morphology, we override the probability-based 
classification and set the class to {\it Galaxy (Star?)}, indicating that 
usually we are looking at misclassified galaxies here, but as a second 
guess we might really be looking at a resolved stellar binary.

The great majority of objects match our templates rather well. Histograms 
of reduced $\chi^2$ values for the best fitting templates show maxima 
around 1.0, just as it is desired statistically (see Fig.~\ref{chi2red}). 
Objects with reduced $\chi^2$ values $>30$ are simply classified as 
{\it Strange Object} whatever their probabilities for different classes 
are. These objects are highly inconsistent with all existing templates 
and many have bad flags anyway. Some of them are clearly real objects 
with accurate photometric data, but unusual spectra, e.g. galaxies with 
extremely strong emission lines or objects with strong variability on a 
time scale of hours or days. For statistical studies of the bulk object 
population we suggest to ignore them, but if your pet object is among 
them, our data could still be useful (depending on the flags).

\begin{figure}
\centering
\includegraphics[clip,angle=270,width=\hsize]{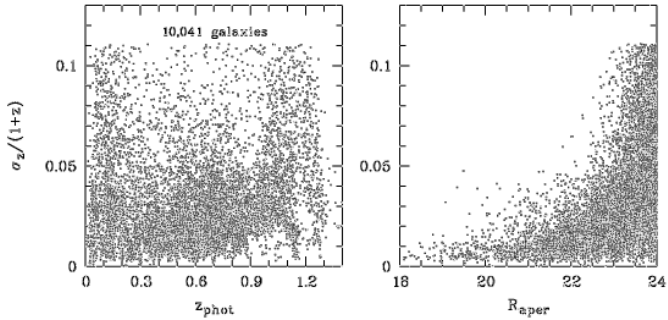}
\caption{Estimated redshift errors: Redshift error vs. redshift (left 
panel) and vs. $R$-band magnitude (right panel) for 10,041 galaxies with 
MEV redshift estimates. The errors are driven by photon noise. At $z\sim 
1$ hardly any galaxies are measured with errors below 0.02, because only 
few of them are sufficiently bright.
\label{szs}}
\end{figure}

\begin{figure}
\centering
\includegraphics[clip,angle=270,width=\hsize]{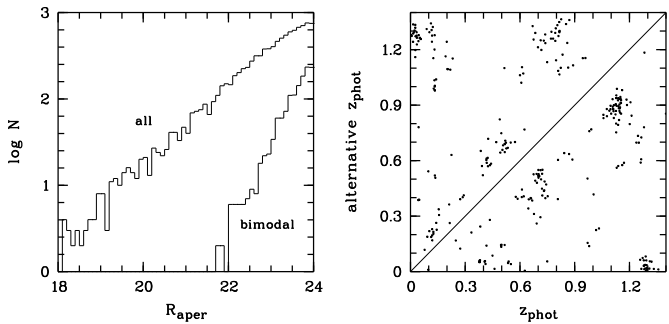}
\caption{Bimodal redshift solutions: {\it Left panel:} $R$-band histogram
of all galaxies with MEV redshift estimates and a subset of those having 
bimodal probability distributions for their photometric redshift. At $23
<R<24$, 20\% of all galaxies have a bimodal $p(z)$.
{\it Right panel:} Alternative (less likely) redshift vs. assigned (more 
likely) redshift for $\sim 1350$ galaxies with bimodal $p(z)$ at $R<24$.
\label{gzabs}}
\end{figure}

\subsection{Photometric redshifts}

As in WMR, we compress the information
of the full redshift probability distribution $p(z)$ of an object into a 
few numbers. These are the actual redshift estimate and its estimated 
error. Such a simplified approach is fine as long as $p(z)$ looks similar
to a Gaussian or is at least not widely distributed. Our estimator of
choice is the {\it Minimum-Error-Variance} (MEV) estimator also known as 
Mean-Square (MS) estimator, which calculates just the mean and variance
of the $p(z)$ distribution. By definition, it minimises the average true
deviation of estimates in a sample. The magnitude dependence of redshift
errors in the galaxy sample (see Fig.~\ref{szs}) demonstrates how these 
errors are driven by photon noise.

Whenever probability distributions are not well represented by a single
Gaussian, a small set of numbers like mean and variance can not convey 
its more complex structure and may be misleading. A common situation in 
photometric redshift determination involves bimodal $p(z)$ functions that
are just the sum of two Gaussians centered at different redshifts. These
cases are the result of ambiguities in colour space, where the filter set
does not discriminate between alternative interpretations. 

Our algorithm detects bimodal cases and compares the probability integral 
under the two modes to decide for the more likely alternative. We then 
list mean and variance of the more likely mode, but we also set a flag
for bimodality and list a second-guess redshift from the other mode. In
the quality saturation domain, the SED is defined clearly enough to avoid
any ambiguities. In the quality transition domain bimodal cases amount to 
4\% of all galaxies at $R=[22,23]$ and 20\% of those at $R=[23,24]$ (see 
Fig.~\ref{gzabs} for a magnitude histogram and the alternative redshifts).

If the $p(z)$ distribution is rather wide, mode decomposition will not 
make much sense anymore, and storing the full distribution would enlarge 
the data volume beyond our intention. Also, the COMBO-17 data confront us 
with this situation only at $R>23.5$, where the classification might be
inaccurate in either case. This is because it faces a number of systematic 
challenges beyond the problem of increased noise, such as our restriction 
in galaxy templates and redshifts and the lack of observed NIR data. For
the same reason, we do not use redshift priors depending on apparent 
magnitude. At $R<23$ such priors are virtually irrelevant compared to the
narrow colour-based $p(z)$ distributions and would not add information. 
At fainter levels they are relevant, but priors could not fix the damage 
caused by all the limitations of restframe UV galaxy templates.

We like to briefly comment on how to deal with wide $p(z)$ distributions 
in hypothetical applications. The MEV estimator will still give the best 
estimate for individual objects if a compression into a single number is      
desired. Of course, for a flat $p(z)$ this will always be the central
redshift of the interval probed, but the average deviation of a sample
will still be minimised to $\sqrt{1/3}$ of the full interval width. 
But plotting a histogram of redshifts thus obtained will be a misleading 
exercise, because after the compression of $p(z)$ into a single number 
any distribution information has been erased. Reconstructing a best 
estimate of the redshift {\it distribution} for all objects combined 
requires summing up the original $p(z)$ distributions.

We decided not to use any MEV estimates, whenever the 1-$\sigma$ error 
exceeded a threshold chosen ad-hoc to be 1/8 of the full redshift interval 
in grid space. This translates to an estimated redshift error threshold
which is constant in $\sigma_z/(1+z)$ due to the changing grid constant 
in redshift space (see also Fig.~\ref{szs}).

\subsection{Known deficiencies}

The classification procedure contains three principal ingredients.
The first one is the {\it data}, largely characterized by the choice of 
filters and exposure times. The second ingredient is the {\it model},
or equivalently our template set. Finally, the third ingredient is the 
{\it classifier}, which may include priors and non-statistical rules.

Our choices on all three ingredients influence the classification quality
on the whole. The filter set leaves room for some class ambiguities, which 
COMBO-17 can not resolve on its own and which lead to misclassifications.
Our simulations of the survey provide us not only with completeness maps,
but also with class cross-contamination maps, which can point to the most
relevant issues (see Sect.~7.3). Also, any template set will always be 
incomplete and can not be expected to catch {\it every} object we come 
across. In the following, we highlight some limitations of the 
classification we are aware of.

{\it Binary stars: }
The stellar library contains no ''M dwarf plus White Dwarf'' binaries,
which have colours that are inconsistent with any single star template 
because of the comparable brightness of the two components. These binaries 
are often serendipitously found in searches for QSOs by optical colour.

{\it Extremely cool stars: }
The stellar library contains no L dwarf spectra, which are not much redder
than M dwarfs in optical colours but have weaker absorption bands. Their 
extremely low luminosity should render them invisible in our sample, and 
the brightest and warmest examples of type L0/1 might still be matched
reasonably well with the Pickles M8 template.

{\it High-redshift galaxies: }
Galaxies at $z>1.3$ show no distinctive features in our filter set and are
typically quite faint. Both factors suggest that these galaxies are mapped
onto a rather broad $p(z)$ distribution, and it should be almost impossible 
to obtain accurate redshift information from COMBO-17. We truncated the
probability determination of galaxy redshifts at $z=1.4$ and deliberately 
excluded higher redshift galaxies from the scope of our work. At this point,
we can not tell what really happens to these galaxies in COMBO-17, because
we are lacking a suitable spectroscopic control sample. In principle, it is 
possible that these objects will contaminate the low-z sample, if a well 
constrained but random match with low-z templates occurs. But it is more 
likely, that $z>1.3$ galaxies are found among those with no MEV redshifts. 

To test the plausibility of this assumption, we now estimate the number of 
expected high-z objects and compare it with the number of objects without
redshifts. Virtually all $z>1.3$ galaxies are at $R>23$, while objects at 
$R>24$ are not within the scope of COMBO-17. Hence, we inspect the interval 
of $R=[23,24]$, where we have a total of $\sim 7300$ galaxies. 
  Baugh \& Efstathiou (1991) provide an approximate redshift
  distribution as a function of median sample redshift. Using this
  result with the median redshift $z_m=0.75$ measured by Brown et
  al. (2003) for COMBO-17 in the interval $R=[23,24]$, we anticipate
  that 10\% of objects in this interval should lie at $z>1.3$. However,
  we find that 17.5\% of objects in this interval have no MEV
  redshifts due to a broad p(z) distribution.
It is hence possible, that all the high-z galaxies hide among those with 
no redshift estimate. However, even if many of the relatively rare high-z 
objects had been mistakenly assigned a low redshift, we believe they would 
not compromise the value of the rich low-z sample.

{\it Seyfert-1 galaxies: }
Seyfert-1 galaxies are detected as QSOs provided their active nucleus is 
sufficiently luminous compared to the stellar light of the host galaxy,
such that the nucleus leaves significant signature in the 17-filter SED. 
Typically, broad-line AGN brighter than $M_B=-21.7$ are identified as AGN
and are tagged {\it QSO}. Fainter broad-line objects are only sometimes
recognized as {\it QSO}, but mostly just classified as {\it Galaxy}. 
At $z>1$ the redshifts of AGN classified as {\it Galaxy} are unreliable,
and the full catalogue could contain up to a few dozen such objects.

{\it Seyfert-2 galaxies: }
Many galaxies belong to the Sy-2 class and can be identified as such in 
deep X-ray observations, but the wavelength resolution of COMBO-17 does 
not allow to tell the difference to normal galaxies, because we can not
clearly see the emission lines. Hence, the AGN nature of these galaxies 
is usually not discovered by COMBO-17, while their redshifts should be as 
reliable as those of normal galaxies.

{\it Compact low-z galaxies: }
At $z<0.2$ and fainter magnitudes, where photometry is less accurate, some 
galaxies show very similar colours as stars even in 16-D colour space. We
break this degeneracy using morphological information. However, very faint 
and unresolved low-z galaxies could still be mistaken as stars, and slightly 
resolved binaries could then be misclassified as galaxies. At present we
have no idea, how common this mistakes could be, but we estimate that at
most it should be a few dozen objects at $R<24$.

\subsection{Derived restframe properties}

The restframe luminosity of all galaxies and quasars are measured from the 
individual 17-filter spectra. For the galaxies, ten restframe passbands are 
considered, SDSS $ugr$ bands, $UBV$ bands in Johnson and Bessell systems and 
a synthetic UV 
continuum band centered at $\lambda_\mathrm{rest} = 280$~nm with 40~nm FWHM 
and a top-hat transmission function. For quasars, we give luminosities in a 
synthetic rectangular passband at 140~nm--150~nm. Depending on the restframe
band and the redshift in consideration, these luminosities are sometimes 
based on extrapolations beyond our filter set. In these cases, the values
as well as their estimated errors tend to be uncertain. At higher redshift,
our filter set probes the restframe UV spectrum which is dominated by light
from the youngest population in a galaxy. Older populations could be hidden
in the restframe UV signal, and only contribute to restframe visual light.
These would not be constrained by the observed SED, but could strongly 
affect the visual restframe passbands.

We derive restframe luminosities by placing the redshifted template that 
corresponds to the galaxy SED into the observed 17-band photometry.
Then we integrate the template spectrum under redshifted versions of the 
restframe passbands. Our reddest restframe band, the SDSS $r$-band, requires 
extrapolation at all redshifts $z>0.5$. The most ultraviolet band, 280/40,
is extrapolated at low redshifts of $z<0.25$. Tab.~\ref{columns} lists
all extrapolation-free redshift ranges. For quasars, we directly measure 
restframe luminosities over $1.4<z<5$ without any extrapolation. 
In Tab.~\ref{rfvega} we give reference data for all restframe passbands.
We would like to ask the user not to trust restframe colours for galaxies 
at $z>1.1$ too much, because there (a) most passbands are extrapolated and 
(b) increasing errors in redshift propagate into the restframe quantities. 
We have not modelled the errors introduced by either effect. 

At least, we have estimated errors for the luminosities from the errors of
the observed photometry. These should be reasonable error estimates in the
non-extrapolation regimes. We add the following three error components in 
quadrature: (i) the magnitude error of the broad-band filter closest to the 
redshifted restframe passband, because it determines the local flux level 
in the passband relative to the whole SED; (ii) the magnitude error of the
total magnitude MAG-BEST, because it determines the overall flux level for
the whole object; and (iii) a minimum error of $0\fm1$ to take into account 
various contributions from redshift errors or from the overall calibration.

We would like to remind the reader of two additional, important sources
of error for luminosities which we have not taken into account. Firstly, 
whenever redshifted restframe passbands lie outside the observed filter set, 
the luminosity estimate relies on an extrapolation of the SED as it has been
fit within the filter set. But true SEDs can deviate from best-fitting
ones, e.g. when an underlying older population shows up in the observed
NIR regime, but is invisible in the observed visual and leaves no trace
in the template then. The reader may be warned that our luminosities are
then only rough estimates. Secondly, large galaxies at low redshifts show
only their central colours in our apertures used for the SED measurement.
Hence, colour gradients inside the galaxy will not be properly reflected in
the galaxy luminosities. Only in the observed-frame $R$-band, the total 
galaxy photometry is correct. All other bands are linked to this measurement
through the SED shape measured in apertures. Colour gradients would affect
all luminosities where restframe passbands are far away from the observed
$R$-band. Throughout the paper, we use $H_0 = h~\times$ 100~km/(s~Mpc) in 
combination with $(\Omega_m,\Omega_\Lambda)=(0.3,0.7)$.

\begin{table*}
\caption{Column entries in the published FITS catalogue. For details of the 
ASCII version, see CDS or COMBO website. 
Some restframe luminosities are extrapolated in some redshift ranges. We
give the redshift intervals, where no extrapolation errors are expected. 
\label{columns}}
\hbox{
\begin{tabular}[t]{ll}
\hline \noalign{\smallskip} \hline \noalign{\smallskip} 
column header  &  meaning  \\ 
\noalign{\smallskip} 
\hline \noalign{\smallskip}
               &  general information \\
\noalign{\smallskip} \hline \noalign{\smallskip}
Nr             &  unique object number \\
ra             &  right ascension (J2000) \\
dec            &  declination (J2000) \\
x              &  x-position on $R$ frame in pixels \\
y              &  y-position on $R$ frame in pixels \\
Rmag           &  total $R$-band magnitude \\
e\_Rmag        &  1-$\sigma$ error of total $R$-band mag \\
Ap\_Rmag       &  aperture $R$-band magnitude in run D \\
ApD\_Rmag      &  difference total to aperture (point sources $\sim 0$) \\
mu\_max        &  central surface brightness in $R$ from SExtractor \\
MajAxis        &  major axis (as observed in $1\arcsec$ seeing) \\
MinAxis        &  minor axis (as observed in $1\arcsec$ seeing) \\
PA             &  position angle, measured West to North \\
phot\_flag     &  photometry flags (see Sect.~3.5) \\
var\_flag      &  variability flag (0 = not variable) \\
stellarity     &  stellarity index from Sextractor \\
\noalign{\smallskip} \hline \noalign{\smallskip}
               &  classification results \\
\noalign{\smallskip} \hline \noalign{\smallskip}
MC\_class       &  multi-colour class (see Tab.~\ref{classtab}) \\
MC\_z           &  mean redshift in distribution $p(z)$ \\
e\_MC\_z        &  standard deviation (1-$\sigma$) in distribution $p(z)$ \\
MC\_z2          &  alternative redshift if $p(z)$ bimodal \\
e\_MC\_z2       &  standard deviation (1-$\sigma$) at alternative redshift \\
MC\_z\_ml       &  peak redshift in distribution $p(z)$ \\
chi2red         &  $\chi^2/N_f$ of best-fitting template \\
dl              &  luminosity distance of MC\_z \\ 
\noalign{\smallskip} \hline \noalign{\smallskip}
               &  total object restframe luminosities \\
\noalign{\smallskip} \hline \noalign{\smallskip}
UjMag           &  $M_{abs,gal}$ in Johnson $U$ (ok at all $z$)\\
e\_UjMag        &  1-$\sigma$ error of $M_{abs,gal}$ in Johnson $U$ \\
BjMag           &  $M_{abs,gal}$ in Johnson $B$ ($z\approx [0.0,1.1]$)\\
e\_BjMag        &  1-$\sigma$ error of $M_{abs,gal}$ in Johnson $B$ \\
VjMag           &  $M_{abs,gal}$ in Johnson $V$ ($z\approx [0.0,0.7]$)\\
e\_VjMag        &  1-$\sigma$ error of $M_{abs,gal}$ in Johnson $V$ \\
usMag           &  $M_{abs,gal}$ in SDSS $u$ (ok at all $z$)\\
e\_usMag        &  1-$\sigma$ error of $M_{abs,gal}$ in SDSS $u$ \\
gsMag           &  $M_{abs,gal}$ in SDSS $g$ ($z\approx [0.0,1.0]$)\\
e\_gsMag        &  1-$\sigma$ error of $M_{abs,gal}$ in SDSS $g$ \\
rsMag           &  $M_{abs,gal}$ in SDSS $r$ ($z\approx [0.0,0.5]$)\\
e\_rsMag        &  1-$\sigma$ error of $M_{abs,gal}$ in SDSS $r$ \\
UbMag           &  $M_{abs,gal}$ in Bessell $U$ (ok at all $z$)\\
e\_UbMag        &  1-$\sigma$ error of $M_{abs,gal}$ in Bessell $U$ \\
BbMag           &  $M_{abs,gal}$ in Bessell $B$ ($z\approx [0.0,1.1]$)\\
e\_BbMag        &  1-$\sigma$ error of $M_{abs,gal}$ in Bessell $B$ \\
VbMag           &  $M_{abs,gal}$ in Bessell $V$ ($z\approx [0.0,0.7]$)\\
e\_VbMag        &  1-$\sigma$ error of $M_{abs,gal}$ in Bessell $V$ \\
S280Mag         &  $M_{abs,gal}$ in 280/40 ($z\approx [0.25,1.3]$)\\
e\_S280Mag      &  1-$\sigma$ error of $M_{abs,gal}$ in 280/40 \\
S145Mag         &  $M_{abs,QSO}$ in 145/10 ($z\approx [1.4,5.2]$)\\
e\_S145Mag      &  1-$\sigma$ error of $M_{abs,QSO}$ in 145/10 \\
\noalign{\smallskip} \hline \noalign{\smallskip}
                &  observed seeing-adaptive aperture fluxes \\
\noalign{\smallskip} \hline \noalign{\smallskip}
W420F\_E        &  photon flux in filter 420 in run E \\
e\_W420F\_E     &  photon flux error in 420/E  \\
W462F\_E        &  photon flux in filter 462 in run E \\
e\_W462F\_E     &  photon flux error in 462/E  \\
\noalign{\smallskip} \hline
\end{tabular}
\begin{tabular}[t]{ll}
\hline \noalign{\smallskip} \hline \noalign{\smallskip} 
column header  &  meaning  \\ 
\noalign{\smallskip} 
\hline \noalign{\smallskip}
                &  observed seeing-adaptive aperture fluxes (cont.) \\
\noalign{\smallskip} \hline \noalign{\smallskip}
W485F\_D        &  photon flux in filter 485 in run D \\
e\_W485F\_D     &  1-$\sigma$ photon flux error in 485/D  \\
W518F\_E        &  photon flux in filter 518 in run E \\
e\_W518F\_E     &  1-$\sigma$ photon flux error in 518/E  \\
W571F\_D        &  photon flux in filter 571 in run D \\
e\_W571F\_D     &  1-$\sigma$ photon flux error in 571/D  \\
W571F\_E        &  photon flux in filter 571 in run E \\
e\_W571F\_E     &  1-$\sigma$ photon flux error in 571/E  \\
W571F\_E        &  photon flux in filter 571 combined \\
e\_W571F\_E     &  1-$\sigma$ photon flux error in 571/combined  \\
W604F\_E        &  photon flux in filter 604 in run E \\
e\_W604F\_E     &  1-$\sigma$ photon flux error in 604/E  \\
W646F\_E        &  photon flux in filter 646 in run D \\
e\_W646F\_E     &  1-$\sigma$ photon flux error in 646/D  \\
W696F\_E        &  photon flux in filter 696 in run E \\
e\_W696F\_E     &  1-$\sigma$ photon flux error in 696/E  \\
W753F\_E        &  photon flux in filter 753 in run E \\
e\_W753F\_E     &  1-$\sigma$ photon flux error in 753/E  \\
W815F\_E        &  photon flux in filter 815 in run E \\
e\_W815F\_E     &  1-$\sigma$ photon flux error in 815/E  \\
W815F\_G        &  photon flux in filter 815 in run G \\
e\_W815F\_G     &  1-$\sigma$ photon flux error in 815/G  \\
W815F\_S        &  photon flux in filter 815 combined \\
e\_W815F\_S     &  1-$\sigma$ photon flux error in 815/combined  \\
W856F\_D        &  photon flux in filter 856 in run D \\
e\_W856F\_D     &  1-$\sigma$ photon flux error in 856/D  \\
W914F\_E        &  photon flux in filter 914 in run E \\
e\_W914F\_E     &  1-$\sigma$ photon flux error in 914/E  \\
UF\_F           &  photon flux in filter U in run F \\
e\_UF\_F        &  1-$\sigma$ photon flux error in U/F  \\
UF\_G           &  photon flux in filter U in run G \\
e\_UF\_G        &  1-$\sigma$ photon flux error in U/G  \\
UF\_S           &  photon flux in filter U combined \\
e\_UF\_S        &  1-$\sigma$ photon flux error in U/combined  \\
BF\_D           &  photon flux in filter B in run D \\
e\_BF\_D        &  1-$\sigma$ photon flux error in B/D  \\
BF\_F           &  photon flux in filter B in run F \\
e\_BF\_F        &  1-$\sigma$ photon flux error in B/F  \\
BF\_S           &  photon flux in filter B combined \\
e\_BF\_S        &  1-$\sigma$ photon flux error in B/combined  \\
VF\_D           &  photon flux in filter V in run D \\
e\_VF\_D        &  1-$\sigma$ photon flux error in V/D  \\
RF\_D           &  photon flux in filter R in run D \\
e\_RF\_D        &  1-$\sigma$ photon flux error in R/D  \\
RF\_E           &  photon flux in filter R in run E \\
e\_RF\_E        &  1-$\sigma$ photon flux error in R/E  \\
RF\_F           &  photon flux in filter R in run F \\
e\_RF\_F        &  1-$\sigma$ photon flux error in R/F  \\
RF\_G           &  photon flux in filter R in run G \\
e\_RF\_G        &  1-$\sigma$ photon flux error in R/G  \\
RF\_S           &  photon flux in filter R combined \\
e\_RF\_S        &  1-$\sigma$ photon flux error in R/combined  \\
IF\_D           &  photon flux in filter I in run D \\
e\_IF\_D        &  1-$\sigma$ photon flux error in I/D  \\
\noalign{\smallskip} \hline
\end{tabular}
}
\end{table*}

\section{Published data}

The published data package includes the object catalogue and coadded images
in the five broad bands of COMBO-17. They are all available from CDS or at
the COMBO-17 website (http://www.mpia.de/COMBO/combo\_index.html). From the
latter site the catalogue is available in both FITS and ASCII format. Given 
the large number of data columns the FITS version is probably the most trivial
one to use in practice. However, we provide an ASCII table for users who 
can not read FITS tables (see the COMBO-17 website for notes on how to read
FITS tables most easily in IDL or MIDAS).

The catalogue lists identifiers, positions, magnitudes, morphologies, as well 
as classification and redshift information as detailed in Tab.~\ref{columns}.
All magnitudes given in the catalogue and this paper use a spectrum of Vega
as their zeropoint, even for filters traditionally defined in an AB system 
(e.g., SDSS $u$, $g$, and $r$). This is also true for all the restframe 
luminosities. If
readers prefer AB magnitudes, a conversion has to be made based on Vega values
given in Tab.~\ref{filterset} and Tab.~\ref{rfvega}. Magnitude values such as 
the total apparent $R$-band magnitude $Rmag$ and all luminosities are 
total object magnitudes based on the SExtractor definition of $MAG\_BEST$.
In contrast, the observed filter flux values are all seeing-adaptive aperture 
fluxes and are calibrated such that they are equal to total fluxes for point 
sources. An exception among the magnitudes is the $R$-band aperture magnitude 
from run D, $Ap\_Rmag$, which is just the Vega magnitude corresponding to
the flux $RF\_D$. Fluxes are given as photon fluxes $F_\mathrm{phot}$ in
units of photons/m$^2$/s/nm. Photon fluxes are related to other flux 
definitions by
\begin{equation}
   \nu F_\nu = hcF_{\rm phot} = \lambda F_\lambda   ~ .
\end{equation}

Photon fluxes are rather practical units at the depth of current surveys.
A magnitude of $V=20.0$ corresponds to 1~photon/m$^2$/s/nm in all systems
(AB, Vega, ST), if $V$ is centered on 548~nm. When the band flux of an 
object is missing, it means that every exposure was saturated in this band. 

The column $ApD\_Rmag$ contains the magnitude difference between the
total object photometry and the point-source calibrated, seeing-adaptive 
aperture photometry:
\begin{equation}
   ApD\_Rmag = Rmag - Ap\_Rmag   ~ .
\end{equation}

On average, this value is by calibration zero for point sources, and 
becomes more and more negative for extended sources of increasing size. 

The $var\_flag$ column reports the number of flux comparisons between 
multiple observations of identical bands, where an object has been found 
as variable. This number is 0 for objects with no such variability, and 
up to 8 for objects which have been found as variable in all the eight 
available comparisons.

The coadded images are provided as FITS images and have slightly different 
coordinate systems. The $U$-band image is rotated with respect to the four 
others, because in February 2000 the WFI camera was serviced and mounted 
back to the telescope in a slightly different orientation. Hence the pixel
coordinates of any given object can differ on the $U$ image by up to five
pixels from the other bands.  The images are $7951\times7595$ pixels in 
size and cover the area common to the $BVRI$ frames. The $U$-band frame has 
the same format although 115 columns ($<30\arcsec$) at the left edge are 
blank due to pointing differences. The instensity levels of the images are 
given in units of photons hitting the detector, so the CCD gain is taken out. 
We strongly recommend readers to use the known spectrophotometric standard 
star for calibration in case they obtain their own independent photometry.

\begin{table}
\caption{Definition of entries for the mc\_class column.   \label{classtab}}
\begin{tabular}{ll}
\hline \noalign{\smallskip} \hline \noalign{\smallskip} 
class entry      &  meaning \\
\noalign{\smallskip} \hline \noalign{\smallskip}
Star             &  stars \\
                 &  (colour of star, stellar shape) \\
WDwarf           &  WD/BHB/sdB star \\
                 &  (colour of WD/BHB/sdB, stellar shape) \\
Galaxy           &  galaxies \\
                 &  (colour of galaxy, shape irrelevant) \\
Galaxy  (Star?)  &  most likely galaxy at $z<0.15$ \\
                 &  (but overlap in colour space with stars) \\
Galaxy  (Uncl!)  &  colour undecided \\
                 &  (statistically almost always a galaxy) \\
QSO              &  QSOs \\
                 &  (colour of QSO, stellar shape) \\
QSO     (Gal?)   &  colour of QSOs, extended shape \\
                 &  (usually Seyfert, but maybe contaminating galaxy) \\
Strange Object   &  very strange spectrum \\
                 &  (unusual intrinsic spectrum, strong photometric \\
                 &  artifacts or uncorrected strong variability) \\
\noalign{\smallskip} \hline
\end{tabular}
\end{table}

\begin{table}
\caption{The restframe passbands and their characteristics. 
\label{rfvega} }
\begin{tabular}{ll|cc}
\hline \noalign{\smallskip} \hline \noalign{\smallskip} 
  name  &  $\lambda_\mathrm{\mathrm{cen}}$/fwhm  & 
  mag of Vega  &  $F_\mathrm{phot}$ of Vega \\
     &  (nm)  &  (AB mags)  &  $(10^8~\mathrm{phot/m^2/nm/s})$  \\ 
\noalign{\smallskip} \hline \noalign{\smallskip} 
(synthetic)     &  145/10  & $+2.33$ & 0.447\\ 
(synthetic)     &  280/40  & $+1.43$ & 0.529\\ 
\noalign{\smallskip} \hline \noalign{\smallskip} 
Johnson $U$     &  365/52  & $+0.65$ & 0.820\\ 
Johnson $B$     &  445/101 & $-0.13$ & 1.407\\ 
Johnson $V$     &  550/83  & $+0.00$ & 1.012\\ 
\noalign{\smallskip} \hline \noalign{\smallskip} 
SDSS $u$        &  358/56  & $+0.84$ & 0.704\\ 
SDSS $g$        &  473/127 & $-0.11$ & 1.305\\
SDSS $r$        &  620/115 & $+0.14$ & 0.787\\ 
\noalign{\smallskip} \hline \noalign{\smallskip} 
Bessell $U$     &  361/65  & $+0.71$ & 0.780\\ 
Bessell $B$     &  441/95  & $-0.13$ & 1.415\\ 
Bessell $V$     &  551/85  & $+0.00$ & 1.009\\ 
\noalign{\smallskip} \hline
\end{tabular}
\end{table}

\begin{figure}
\centering
\hbox{
\includegraphics[clip,angle=270,width=0.5\hsize]{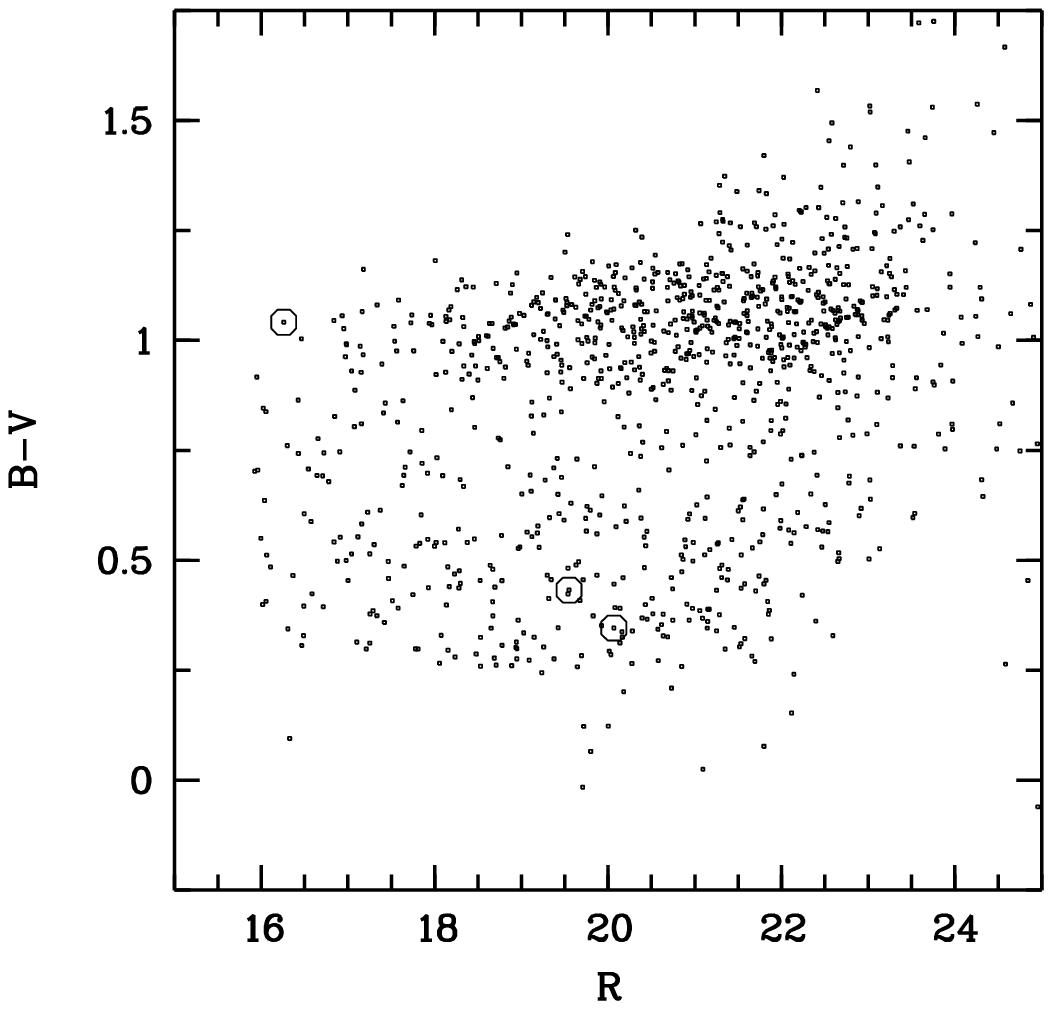}
\includegraphics[clip,angle=270,width=0.5\hsize]{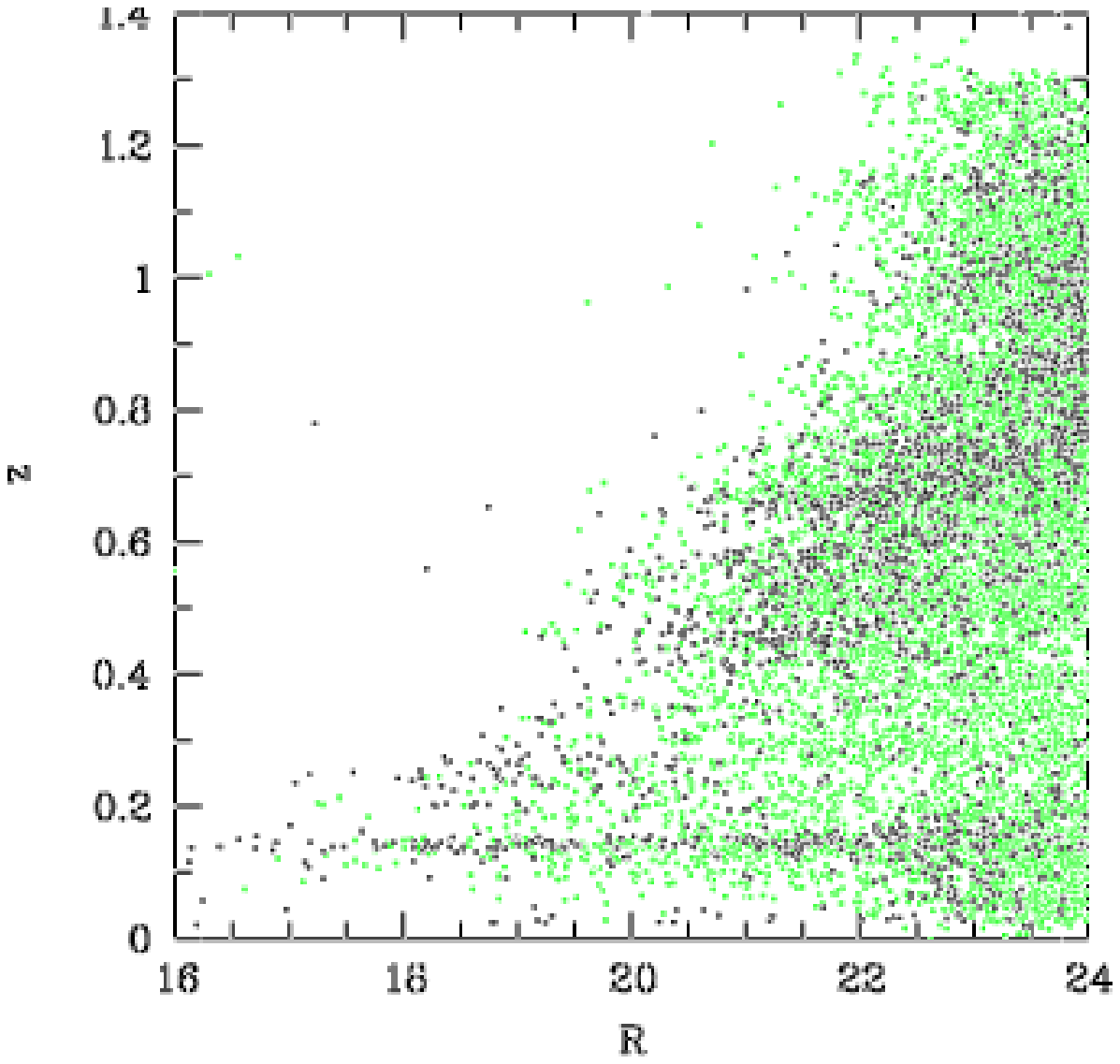}}
\caption{{\it Left panel:} Star sample: $B-V$ colour vs. $R$-band 
magnitude. Variable stars are marked by a circle.
{\it Right panel:} Galaxy sample: MEV redshift estimate vs. $R$-band 
magnitude ({\it black:} red sequence galaxies, {\it grey:}
star-forming galaxies).
\label{starsample}}
\end{figure}

\begin{figure*}
\centering
\includegraphics[clip,angle=270,width=15cm]{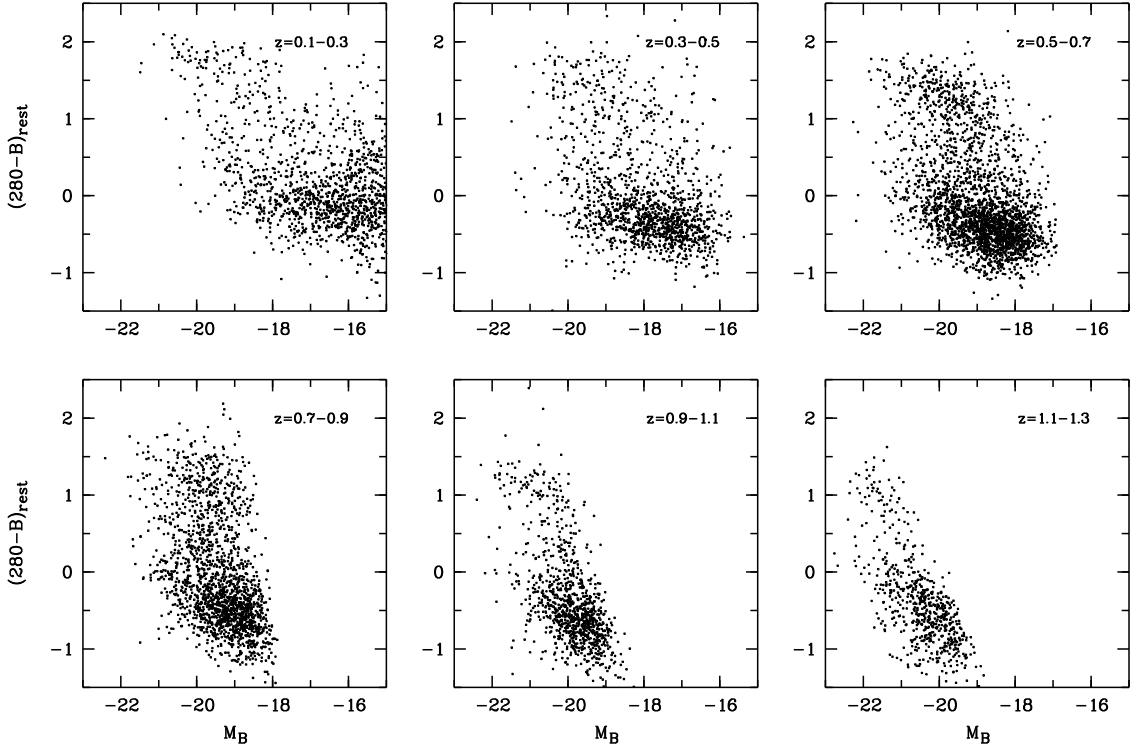}
\caption{Galaxy sample: Restframe $280-B$ colour vs. $M_B$ luminosity. 
The sharp sample cutoff on the right corresponds to $R=24$, the most
reliable subsample. The distribution is clearly bimodal, with a
red sequence and a blue cloud of star-forming galaxies.
\label{gal28mb}}
\end{figure*}

\begin{figure}
\centering
\includegraphics[clip,angle=270,width=\hsize]{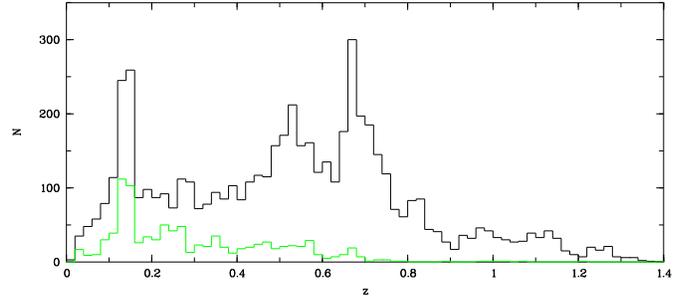}
\caption{Galaxy sample: Redshift histogram from MEV estimates at 
magnitude limits of $R<23$ (black line) and $R<21$ (grey line). 
The redshift distribution in the CDFS is clearly unusual and not
representative of the cosmic average. The redshifts themselves
are quite reliable as the comparison in Sect.~8 shows.
\label{galzhist}}
\end{figure}

\begin{figure}
\centering
\includegraphics[clip,angle=270,width=\hsize]{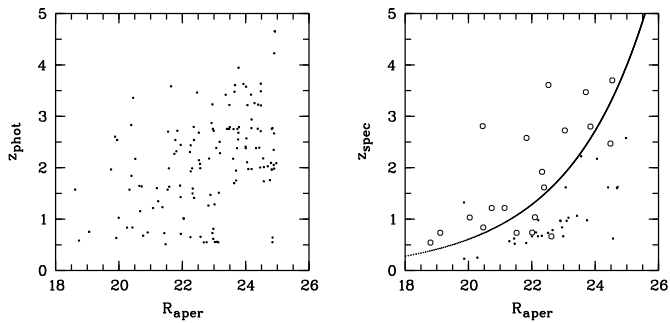}
\caption{Quasar sample: {\it Left panel:} Redshift vs. $R$-band magnitude 
for all identified quasars with MEV redshift estimates.
{\it Right panel:} The purely optical COMBO-17 selection of type-1 AGN is
basically complete at luminosities $M_{B,{\rm Vega}}<-21.7$ (solid line), 
where nuclear light contributes significantly to the SED. At lower AGN 
luminosities the SED is dominated by stellar light from the host galaxy.
Shown here are all type-1 AGN identified in X-ray follow-up of the CDFS 
1~Msec image. Large circles mark those identified in COMBO-17 as quasars, 
and small dots those classified as normal galaxies -- their AGN nature 
can only be discovered with X-rays or spectroscopy.
\label{qsosample}}
\end{figure}

\section{Classified object samples} 

Most work done with the catalogue will either be the selection of samples 
according to some criteria, e.g. 'all stars', 'all QSOs at $z>2$' or be the
identification of sources with known position, e.g. X-ray counterparts. 
However, for the characterization of the data we present some figures which
can immediately be plotted from the published catalogue. We give some sample
queries below, where we also use wildcards ({\it *}) when selecting objects
by class name. These queries will mostly address the magnitude range from
$R=16$ to $R=24$ in terms of the aperture magnitude $Ap\_Rmag$. Brighter 
than $R=16$ some flux measurements are saturated and the SED is potentially
incomplete. Faintwards of $R=24$ the completeness and reliability of the
redshifts and classifications drop too far down.

\begin{figure}
\centering
\includegraphics[clip,angle=270,width=\hsize]{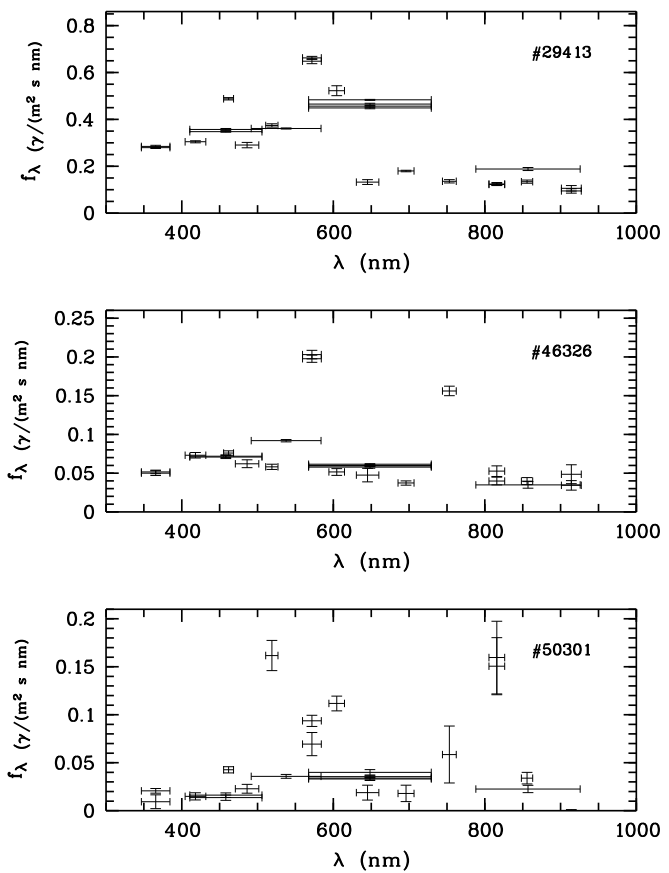}
\caption{Three objects at R=[16,24] were classified as {\it Strange Object} 
based on their remarkably poor fits to all available templates (see text 
for more details).
\label{strgSEDs}}
\end{figure}

\subsection{Stars and 'WDwarfs'}

A query for $MC\_class='Star*'$ and $16<Ap\_Rmag<24$ selects 997 objects 
from the catalogue. Five of them have bad flags ($phot\_flag>7$), 
but their photometry looks fine to us. At $R<16$, stars 
are saturated in the deep $R$-band stack and get the saturation flag in 
the catalogue. Most filters were observed with additional short exposures 
providing a flux measurement where saturation prevents it (unfortunately 
not the $R$-band in run D). Hence, most aperture photometry (calibrated to
resembe full magnitude for point sources anyway) should be fine when 
measurements appear in the catalogue. The final saturation limits range 
from $m\sim 12$ at the far-red end to $m\sim 15$ at the blue end with 
exception of the $R$-band in run D which lacks short exposures and hence
saturates at $m\sim 16$. Nine more objects are identified as {\it WDwarf}. 
They range within $19<R<22$ and have all fine flags.

Fig.~\ref{starsample} shows a colour-magnitude diagram of the star sample,
including three variable stars and the blue {\it WDwarf}s. We would like
to point out again, that our {\it WDwarf} class contains indeed not only
white dwarfs but all stars bluer than spectral type F. This includes BHB 
(Blue Horizontal Branch) stars, Blue Stragglers and sdB stars. The proper 
white dwarfs are actually only discovered if they are hotter than 6000 K,
but we would expect no cool object in our small fields anyway.

\subsection{Galaxies}

A query for $MC\_class='Galaxy*'$, $phot\_flag<8$ and $16<Ap\_Rmag<24$
selects 11343 objects from the catalogue. A sample of 86 further galaxies 
at $Ap\_Rmag<24$ have bad $phot\_flag \ge 8$. Constraining the sample with
good flags ($<8$) to those with MEV redshifts yields 10046 objects. The 
remaining 1297 galaxies are virtually all fainter than $R=23$, reflecting 
just the incompleteness of MEV redshift estimation at faint magnitudes. 
Some 13000 fainter $R>24$ galaxies have redshifts, but they are of 
decreasing quality and certainly form in no sense a complete sample.

Fig.~\ref{starsample} shows a Hubble diagram of the galaxy sample revealing
some structure in redshift space as well as a few maybe surprisingly bright 
objects. One of them is probably a misclassified star, the nature of the
others is currently unclear. We suggest to exclude them from any sample for
statistical studies.

Fig.~\ref{gal28mb} shows colour-magnitude diagrams of the galaxy sample in
several redshift slices. A clear bimodal distribution with a red-sequence
and a star-fomring blue cloud can be seen at all redshifts.

Fig.~\ref{galzhist} shows the redshift distribution of galaxies with $R<21$ 
or with $R<23$. It deviates from the mean expected distribution due to the
finite size of the field and large-scale structure. A cluster at $z\sim 
0.15$ is clearly visible and the known sheets at $z\sim 0.67$ and $z\sim 
0.73$ \citep{GCD03} are blended into one peak in this histogram.

\subsection{Quasars}

A query for $MC\_class='QSO*'$ and $16<Ap\_Rmag<24$ selects 97 objects. 
They all have MEV redshifts and no bad flags (see Fig.~\ref{qsosample}). 
Fainter QSOs could be selected, but probably with low completeness. We also
inspected the subarea observed by the Chandra X-ray observatory for 1~Msec.
Across all redshifts 16 type-1 AGN with $M_B<-21.7$ were identified by 
spectroscopic follow-up \citep{Szo04}. A comparison reveals that COMBO-17 
missed only one of the 16 type-1 AGN, suggesting a selection completeness 
of type-1 AGN well above 90\% even into the Seyfert-1 luminosity regime
(see Fig.~\ref{qsosample}).

The Deep Field observations of Chandra and XMM pick up almost all QSOs 
found by COMBO-17 in their fields of view. Only six out of 48 QSOs in the 
XMM area are not contained in the XMM source catalogue. Three of these are 
in the Chandra area and are confirmed as AGN with redshifts. A fourth one 
is a very weak, marginal Chandra source, and the remaining two are outside 
of the Chandra area. Hence, we don't know whether Chandra would have 
detected them in 1~Msec observations.

\subsection{Strange objects}

A query for $MC\_class='Strange~Object'$ at $16<Ap\_Rmag<24$ selects three 
objects, none of which has bad flags (see Fig.~\ref{strgSEDs}). The first 
one of these, object 29413, is an isolated point source of $R\sim 21$.
It shows no signs of variability, because the four $R$-band measurements
from different epochs are totally consistent. But it appears quite hard to 
explain its SED without brightness variations. The $R$-band appears much
brighter on the whole than some patches within its broad passband, which 
are probed by the medium-bands redwards of 600~nm. In a non-variable
source this would imply enourmous fluxes between the low medium-bands.
At this stage, we have no explanation for the nature of this object.

The second object (46326) is an extended source of $R\sim 23$ and can 
easily be explained by a galaxy with extremely strong emission lines.
A plausible explanation places this object at $z\sim 0.15$, so it would
show its H$\alpha$/N{\sc ii} lines in the filter 753/18 and its H$\beta
$/O{\sc iii} line in the filters 571/25 and $V$-band. The 753/18-filter 
shows emission at an equivalent width of $\sim55$~nm within its limits 
of transmission, while 571/25 and $V$-band consistently show $\sim50$~nm 
of emission.

The third object (50301) is totally swamped in the scattered light halo
of a superbright ($R\sim 10$) foreground star. It is probably not safe 
to interpret the measured SED for this object.

\subsection{Variable objects}

COMBO-17 is not tailored to detect variability in any complete sense. Still, 
a query for $var\_flag>0$, $phot\_flag<8$ and $16<Ap\_Rmag<24$ selects 175 
variable objects in the CDFS. Four more variables have bad flags. Three of 
them are stars, presumably two RR~Lyrae and one M~star (see 
Fig.~\ref{starsample}). All others are classified as QSOs and galaxies.  

At $R<22$, 34 out of a total of 39 QSOs are observed to be variable
already in the COMBO photometry, while at $22<R<24$ the fraction is 
32 out of 58 QSOs. Hence, the rate of observed variability drops from 
87\% to 55\% due to the increased photometric noise that requires 
higher amplitudes to detect variability.

We note, that within the Chandra 1~Msec many of the variable galaxies 
are detected in X-rays and some of these have spectroscopic IDs as AGN. 
Some variable galaxies are likely to be rendered as such by Supernovae
(see Wolf et al. 2001c for some cases). 

\begin{figure}
\centering
\includegraphics[clip,angle=270,width=\hsize]{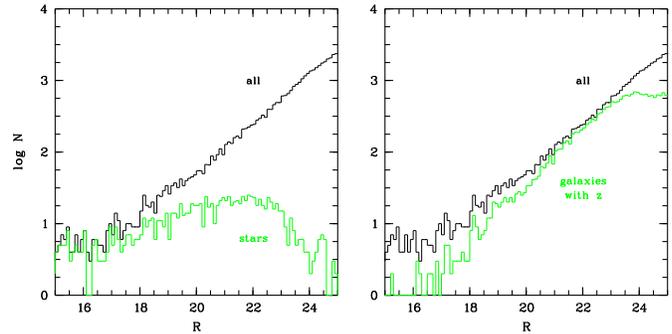}
\caption{Number counts: The black solid line shows all objects in the 
catalogue. The grey line shows the subset of classified stars in the
left panel and the subset of galaxies with MEV redshift estimates in
the right panel. The counts are plotted over the total magnitude $Rmag$. 
\label{ncounts}}
\end{figure}

\begin{figure*}
\centering
\hbox{
\includegraphics[clip,width=0.33\hsize]{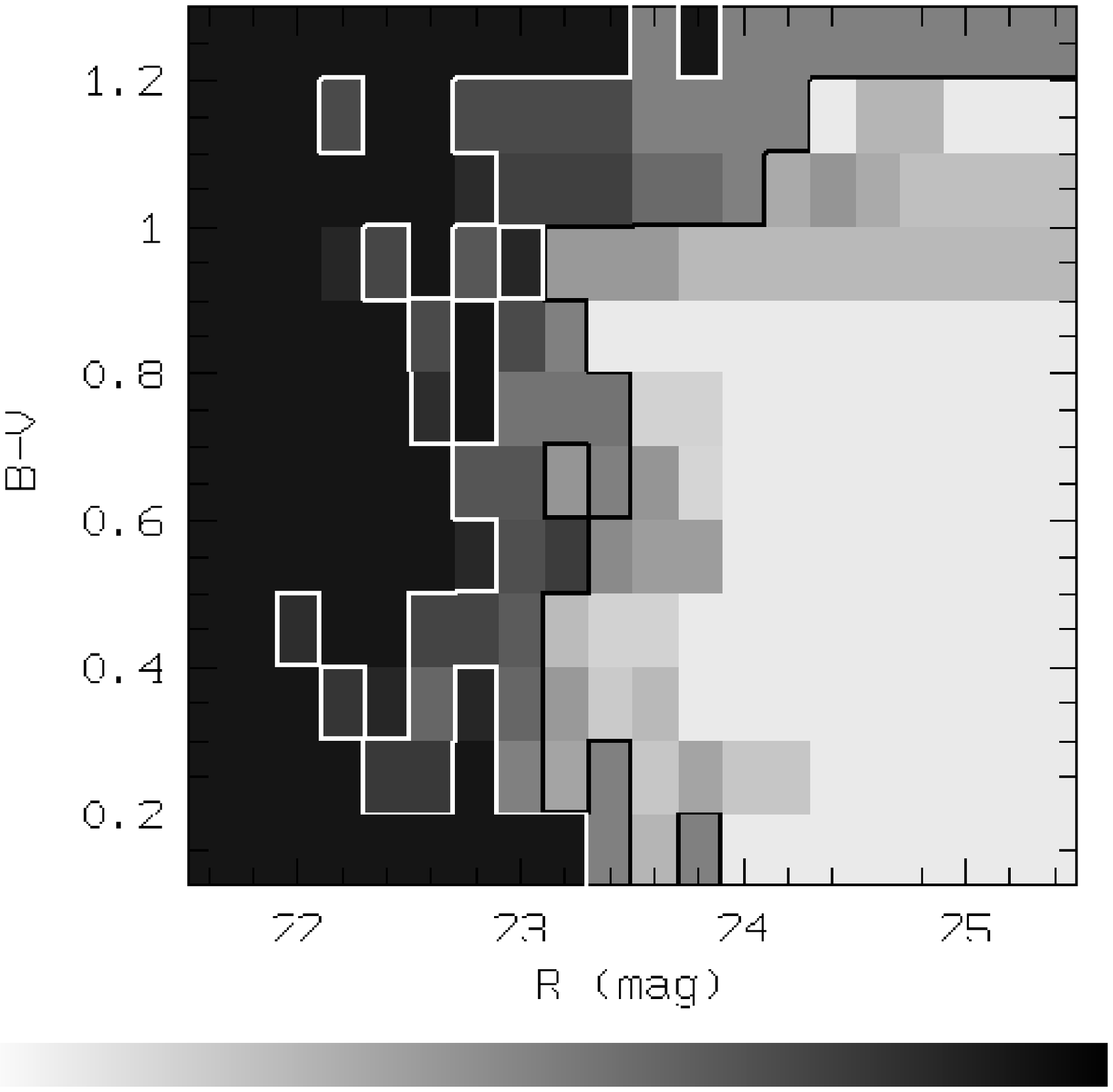}
\includegraphics[clip,width=0.33\hsize]{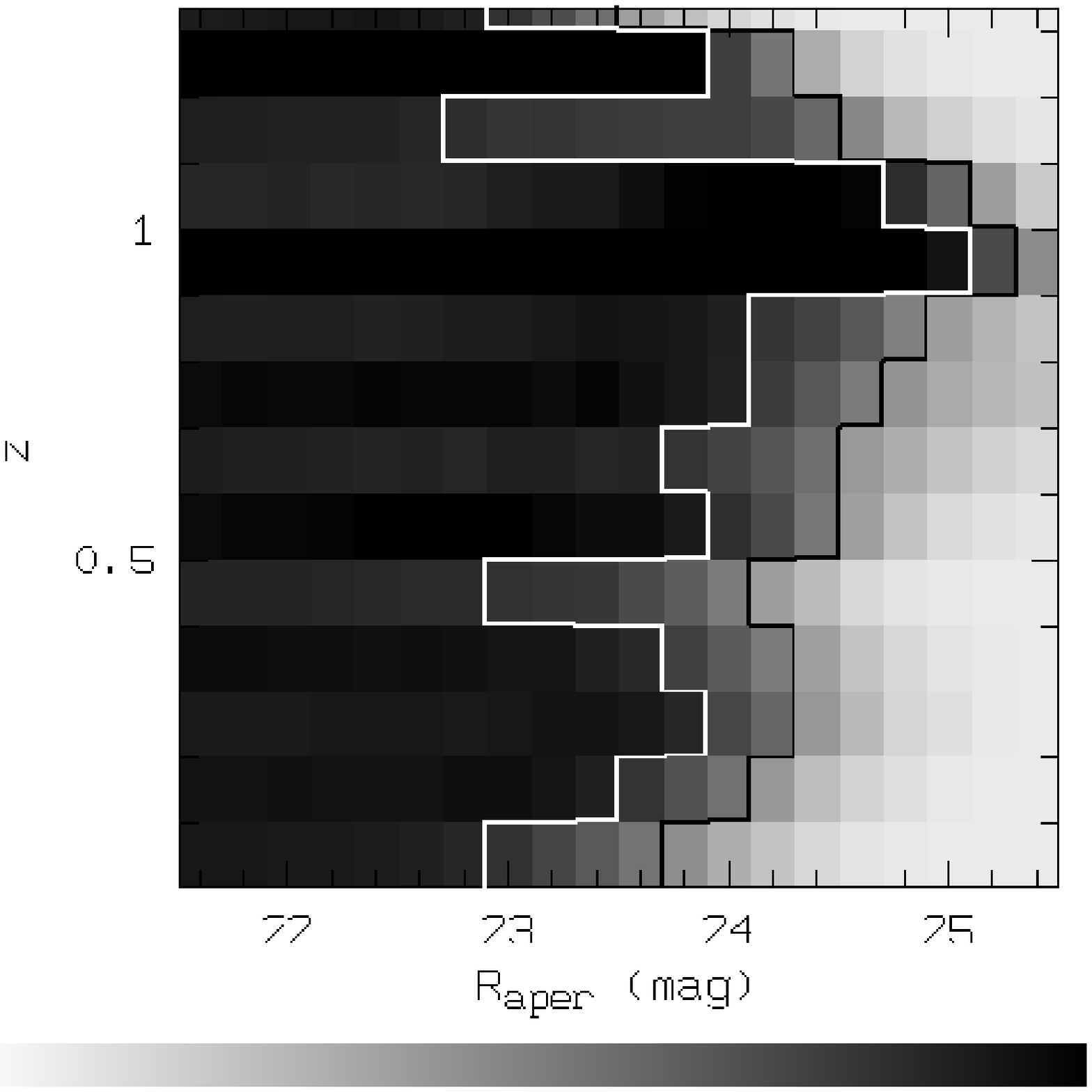}
\includegraphics[clip,width=0.33\hsize]{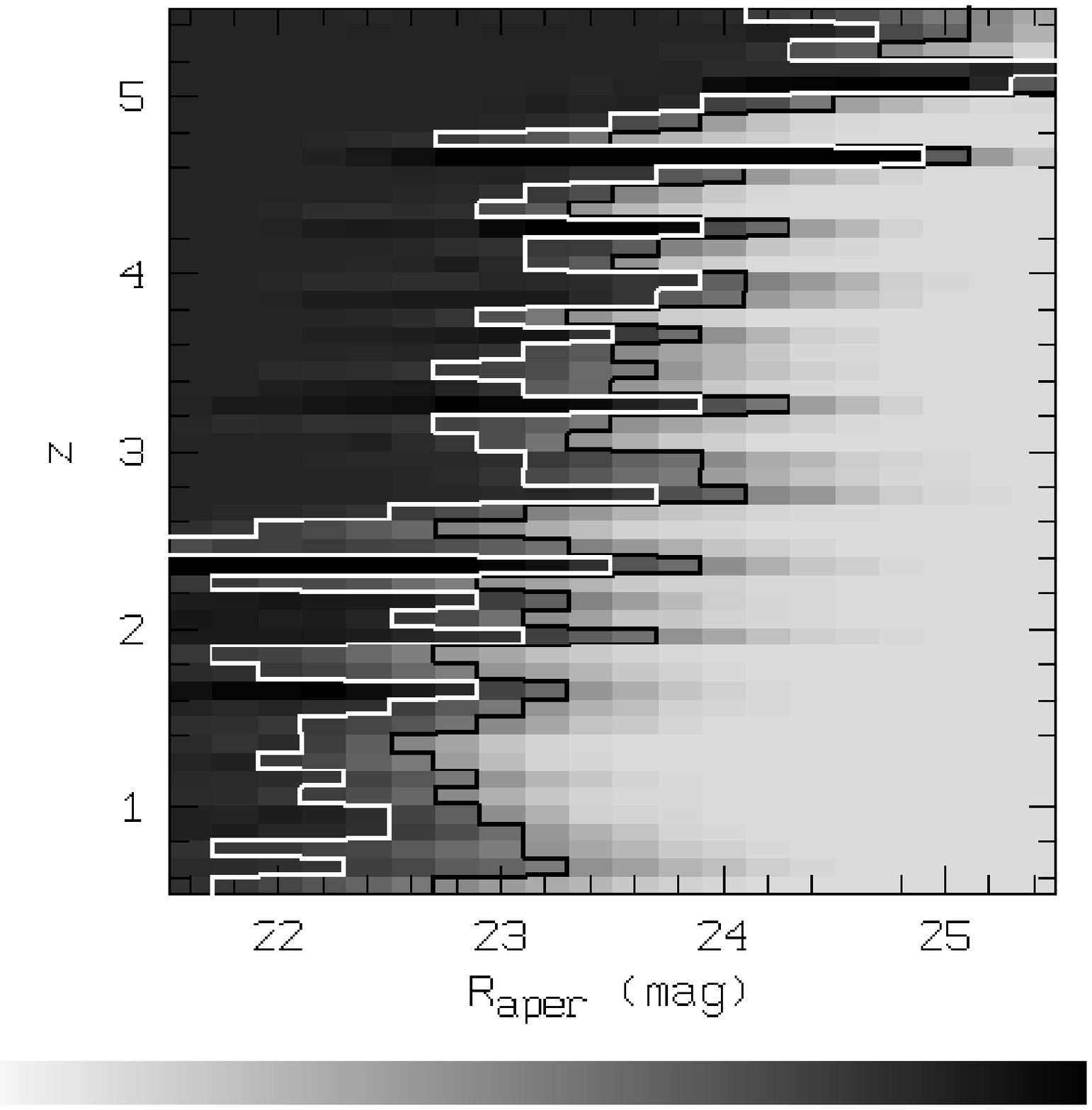}}
\caption{Completeness map: Grey-scale maps of completeness from Monte-Carlo
simulations of the survey (black: 100\%, light grey 0\%).
{\it Left panel:} Stars, $B-V$ colour vs. $R$-band magnitude.
{\it Center panel:} Galaxies, redshift vs. $R$-band magnitude.
{\it Right panel:} Quasars, redshift vs. $R$-band magnitude.
\label{cmapthree}}
\end{figure*}

\section{Simulations of survey performance}

\subsection{Basic technique}

We carried out simulations of the survey in order to map completeness 
and contamination expected for the filter set and imaging depths 
given in COMBO-17. The basic technique was already described in WMR: 
We created lists of test objects from the colour libraries in Sect.\,3 
across a range of aperture magnitudes as in $Ap\_Rmag$. For each object 
in the library we determined filter fluxes and photometric errors. 
Then we scattered the flux values of the objects according to a normal 
distribution of the errors seen in the COMBO-17 observations. Finally,
we recalculate the resulting colour indices and index errors and use 
this object list as an input to the classification. 

The simulations show how well the classification can potentially work, 
assuming that library spectra precisely mimic real objects. In reality, 
differences between assumed and real SEDs will degrade the performance. 
Nevertheless, the simulation highlights the principal shortcomings of 
the method itself and the chosen filter set in particular. 

We run these tests for stars, galaxies and quasars with magnitudes in the
range of $R=[21.6,25.4]$, in order to see how the classification degrades 
from optimum to useless with decreasing object flux. At $R=23$ objects are
measured roughly at photon noise levels of 10-$\sigma$ in the medium-band 
filters. But for bright objects, calibration errors and fine differences 
between real and model SEDs dominate the total errors. Thus, we first add 
a 5\% uncertainty quadratically on top of the colour index errors before 
applying the classification, just as we do it with real catalogue objects.
The performance of the classification converges to its best level at $R\la
22$. At the other end, $R\approx 25$ objects are well detected only in the 
broad-band filters, while the medium bands yield only fluxes with errors 
higher than 40\%. We expect the survey to be almost useless at this level.
We use one realisation of each library at each point in magnitude value.
The latter are spaced in steps of $0\fm2$. This leads to a total of, e.g.,
about 1.3 million galaxies in the simulation.

Finally, we compare the classes and redshifts of the classified output
with those defining the input objects. For any object type and redshift
range, we can quantify the completeness of the classification as the 
fraction of input objects which are recovered correctly. But we can also
look at the rate of objects being misclassified and hence contaminating 
samples of other classes. In the following, we discuss the completeness
and contamination maps in detail.

\subsection{Expected completeness of class samples}

\subsubsection{Stars}

The completeness of star samples depends on their colour and magnitude as 
shown in the map of Fig.~\ref{cmapthree}. The map suggests that for stars 
with $B-V < 1.0$ (roughly spectral types FGK) the completeness does not 
vary much with colour, only with magnitude. The 90\% completeness limit 
of FGK stars is at $R\approx 22.5$ and the 50\% level is reached around 
$R\approx 23.2$. Redder M stars at $B-V > 1.0$ appear much brighter in the 
far-red bands for any given $R$ magnitude. Also, they are among the reddest 
objects of all templates. Both factors help to increase the $R$-band depth 
of their successful classification. The result of these simulations is
confirmed by the observed completeness limits in the stellar sample. The
colour-magnitude diagram of the star sample in Fig.~\ref{starsample} shows
how the M~stars reach deeper than bluer stars. The average completeness
can be better assessed from the number count plot of all objects vs. the
stars in Fig.~\ref{ncounts}.

\subsubsection{Galaxies}

The completeness of galaxy samples not only depends on magnitude and 
redshift (see Fig.~\ref{cmapthree}), but also on the restframe colours of 
the galaxy. This map, however, shows the average completeness for galaxies 
of all colours. The depth of the overall classification increases when 
going from $z=0$ to $z=1$ mostly because red galaxies can be identified to 
greater depth with increasing redshift, as they turn into increasingly 
red objects with increasingly bright far-red fluxes for given $R$-band 
magnitudes (like M stars). For the bluest starburst galaxies, the depth of 
the classification changes hardly with redshift. This is discussed in 
more detail in Wolf et al. (2003a), although actual map details have 
changed following the changeover in templates. Again, the completeness
derived from number count plots (see Fig.~\ref{ncounts}) confirmes the
basic result of the simulations.

\subsubsection{Quasars}

A previous version of the completeness map for quasars in COMBO-17 has 
already been published in a paper on the evolution of the luminosity 
function of quasars \citep{Wolf03b}. However, the recent change in the 
galaxy templates has also affected the quasar classification, and here
we present the updated map for the catalogue published here. The map 
shows oscillations in the depth of the classification. They are caused
by the signature of strong emission lines migrating through the filter
set and changing between strong visibility in a medium-band filter and
invisibility in the spectral blind spots between the medium-band filters.

This completeness map assumes QSO spectra to comprise only AGN light, 
being based on the SDSS template and dominated by high-luminosity QSOs. 
Mixing stellar light into the SEDs should reduce the completeness until 
the AGN nature can not be detected any more. If host galaxies dominate 
the spectrum of an active galaxy, it will not be classified as a QSO but 
as a galaxy. Fig.~\ref{qsosample} compares broad-line AGN identified in 
X-ray follow-up with QSOs identified by COMBO-17 in the same area. It 
suggests, COMBO-17 is $>90$\% complete at picking out type-1 AGN with 
luminosities above $M_B=-21.7$, but starts to fail below that level of
nuclear activity. A consistent explanation involving the simulated map
and the nuclear contrast issue works as follows: at low redshifts the 
real limitation is the nuclear activity level depressing the observed
completeness below the simulated one. At higher redshifts the limit of
$M_B=-21.7$ drops to faint apparent magnitudes where the survey depth 
limits identification.

\begin{figure*}
\centering
\hbox{
\includegraphics[clip,width=0.33\hsize]{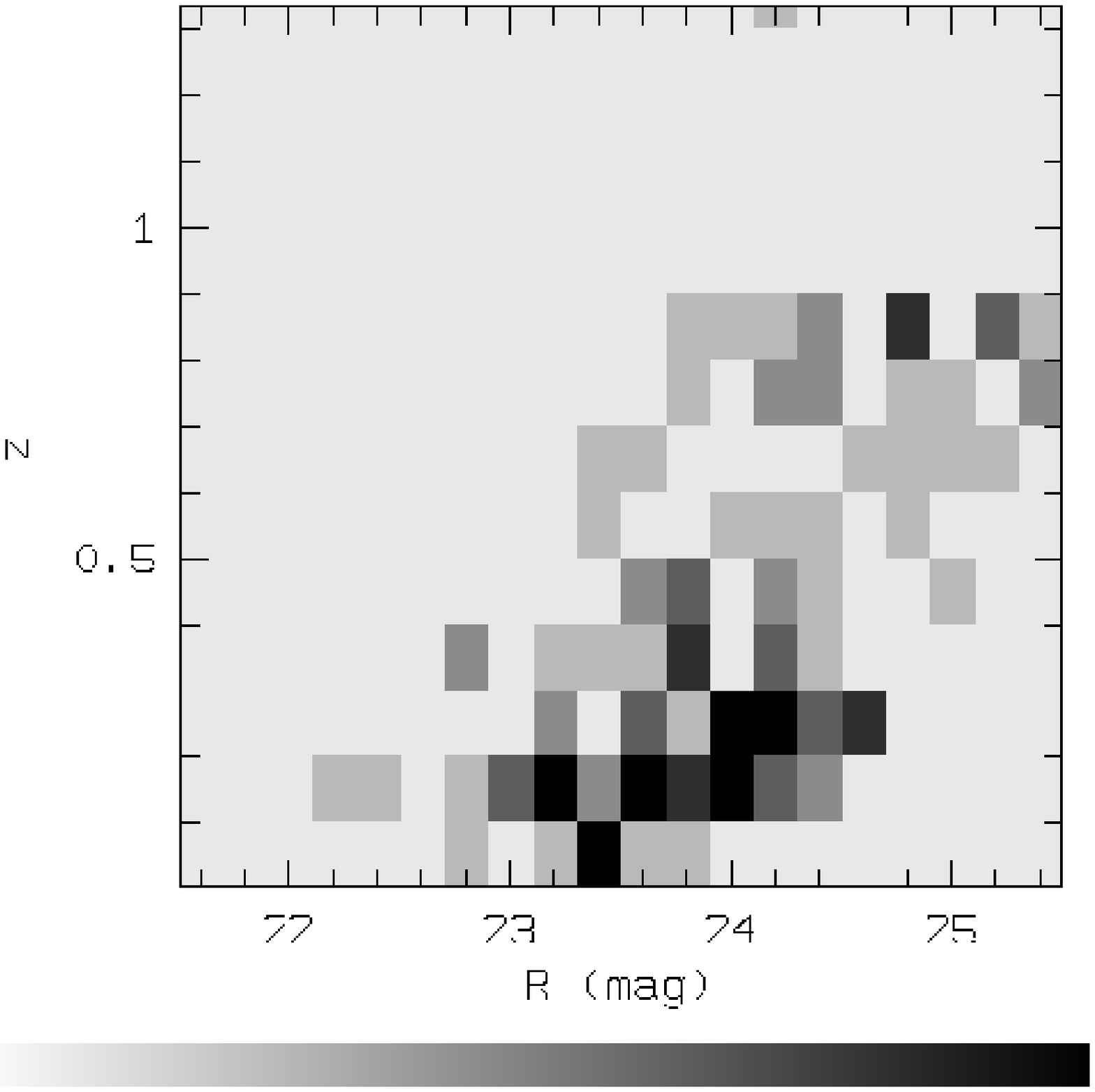}
\includegraphics[clip,width=0.33\hsize]{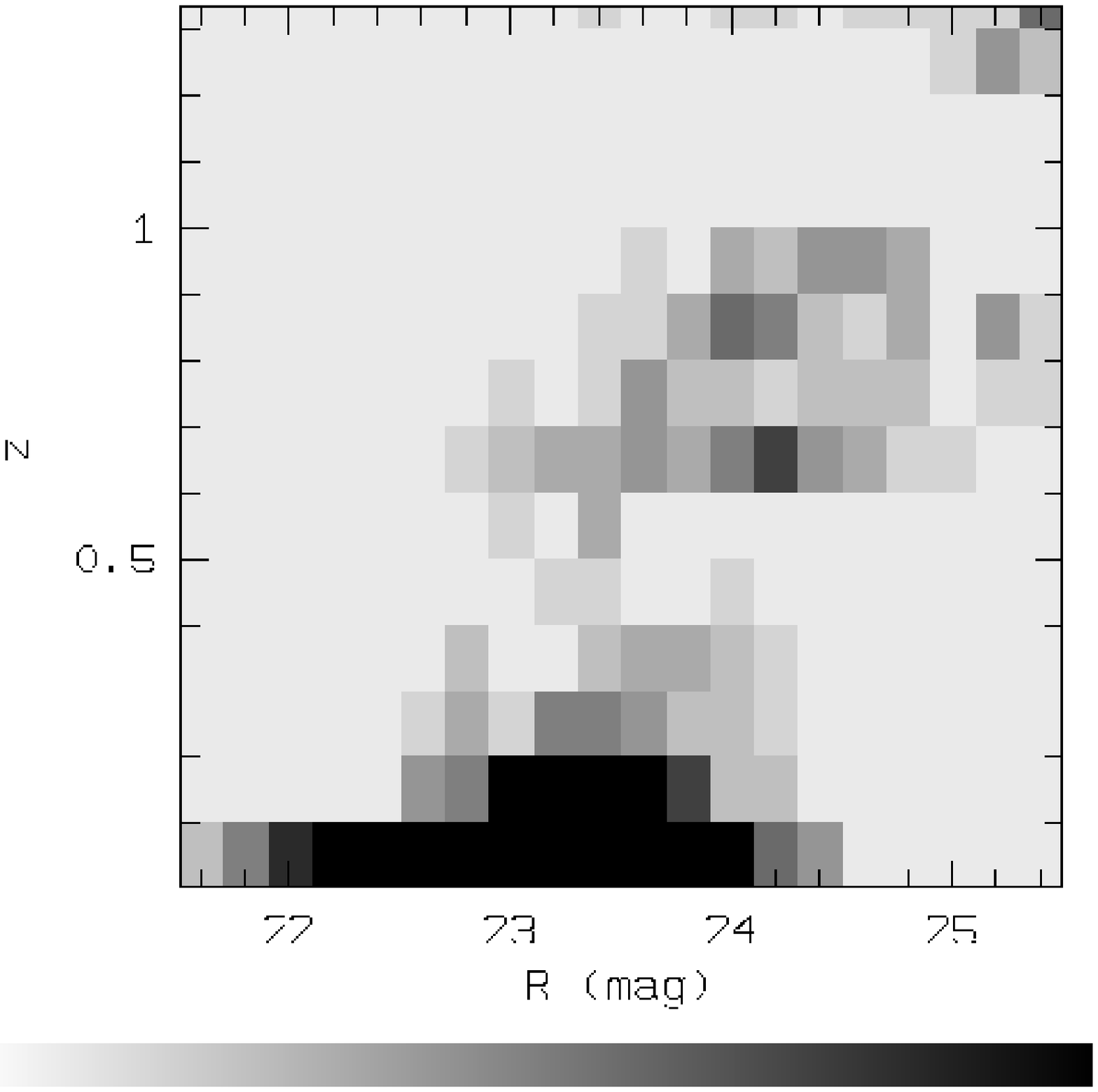}
\includegraphics[clip,width=0.33\hsize]{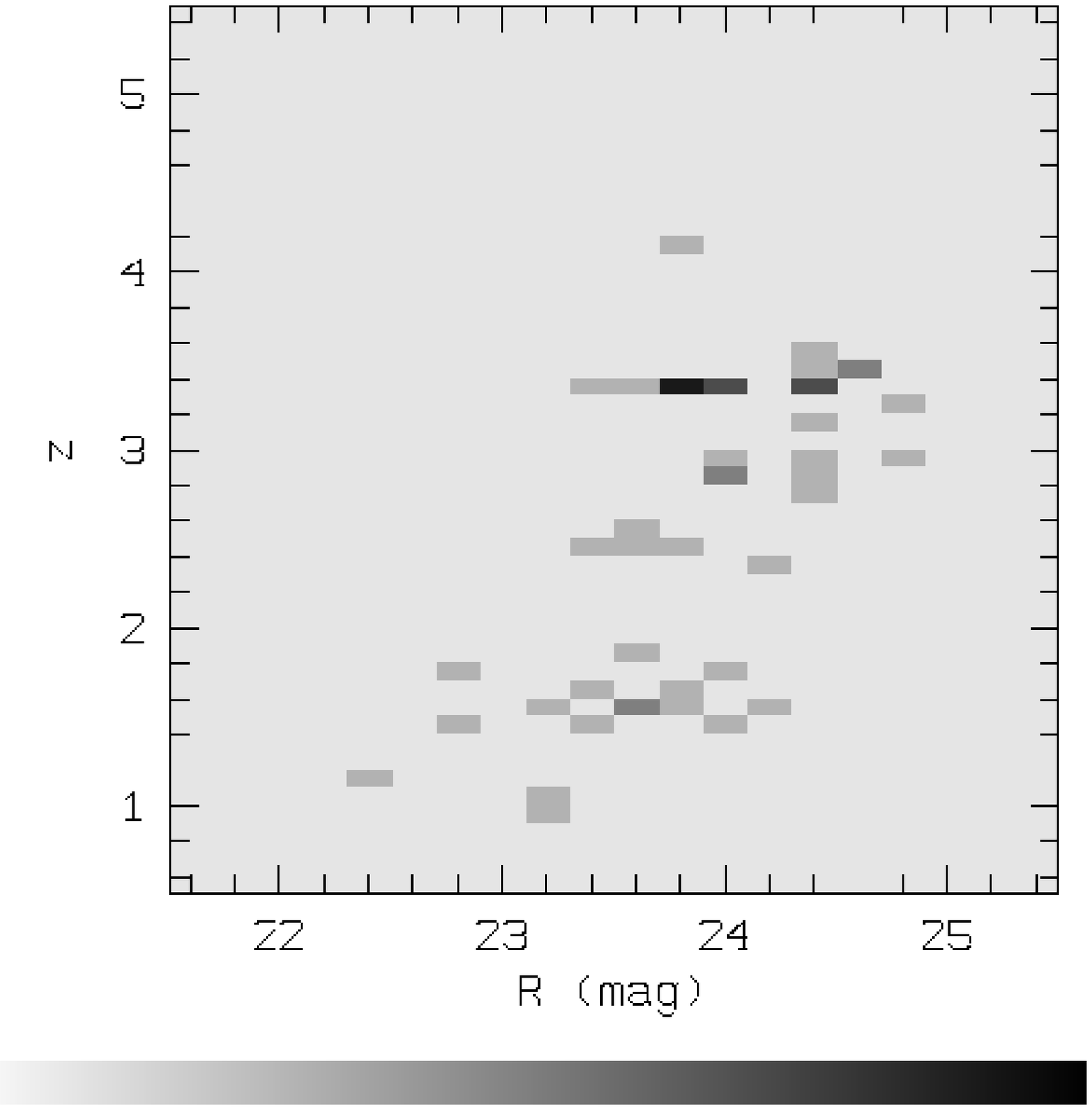}}
\caption{Contamination maps: Grey-scale maps of class cross-contamination 
from Monte-Carlo simulations of the survey (black: more contamination, 
light grey: no contamination, normalisation arbitrary, details see text).
{\it Left panel:} Stars, misclassified as galaxies at redshift $z$ vs. 
$R$-band magnitude. This map basically reflects the incompleteness of
the star selection at faint magnitudes.
{\it Center panel:} Galaxies at redshift $z$ vs. $R$-band magnitude,
which are misclassified as stars. Especially, around zero redshift, 
galaxy SEDs are easily confused with SEDs of stars.
{\it Right panel:} Galaxies, misclassified as quasars at redshift $z$ vs. 
$R$-band magnitude.
\label{Kmapthree}}
\end{figure*}

\subsection{Expected contamination in class samples}

There is a multitude of contamination maps we could possibly wish to plot.
Each of the four classes can lose objects to the other three classes. 
Each of those twelve plots can be shown over two sets of class parameters
--- if you wish to see where you lose objects from a class you plot the
map over input parameters, and if you wish to see where exactly they go,
you plot it over output parameters. Here, we just present the three most
important contamination maps (see Fig.~\ref{Kmapthree}).

The biggest issue in the whole COMBO-17 classification is confusion between 
stars and galaxies. First, all stars not successfully identified as such 
are either classified as {\it Galaxy} or {\it Galaxy (Uncl)}. The chances
of a star being mistaken as a quasar are less than 1 in 1000 as far as the
simulations are concerned (although stars outside our template range pose
a problem, like binaries of M~dwarf$+$white dwarf). So, the incompleteness 
of stars at $R>23$ translates directly into a contamination of the galaxy 
class there. At $R=23.5$, the map suggests that a quarter of all K~stars 
are turned into {\it Galaxy} (not counting {\it Galaxy (Uncl)}), and at 
$R=24$ it is already two thirds. 

How much trouble K~stars mean for any analysis of galaxy samples depends 
on the number counts of stars, and hence on the Galactic coordinates. We
briefly discuss a very rough estimate for the CDFS: We expect about 200 
stars per magnitude bin at $R\sim 23$, of which (roughly guessing) 80 may 
be K stars. In the end, the contamination maps seem to suggest that at  
$R=[23,24]$ around 20 to 40 K~stars are misclassified as {\it Galaxy}.
In ground-based seeing, it will be hard to reliably identify the stars
with a morphological criterion, because some faint galaxies will also be
considered unresolved. A final classification could be settled using the 
HST imaging of GOODS and GEMS.

We note, that M~stars ($B-V>1$) are correctly identified to fainter levels 
than K~stars and do not pose a big problem. This is plausible given that 
they show more flux in far-red filters, securing their identification.

Second, even at brighter levels there is confusion between certain stars and 
galaxies near redshift zero due to their intrinsically similar colours. The 
maps indicate that almost 1\% of the galaxies at $z<0.15$ and $R=[22,24]$ end
up being classified as stars, if we do not override the statistical class
decision by a morphological criterion. The catalogue contains about 750 
galaxies within these limits, ten of which would have been classified as 
stars if it were not for the extended morphology (see Sect.~4.4 and 4.6).

A less important problem are galaxies contaminating the quasar class. In
pure broad-band surveys this is an important issue, but in COMBO-17 it
appears to be less relevant. The simulations suggest there would be a 
1-in-10000 chance that galaxies at $R\sim 24$ are mistaken as quasars.
This would imply one fake quasar at $R<24$ which is really a galaxy. But
the main contamination in the quasar class probably arises from objects
with unusual spectra not represented by the templates, hence there could
be a few objects in total.

The last relevant issue is that of quasars contaminating the galaxy class.
Here, again the completeness maps of quasars suggest that whatever quasar
is not successfully identified, will end up in the galaxy sample. We have
discussed above in detail, that proper quasars of $M_B<-23$ are probably
all correctly identified, while incompleteness in the detection of fainter
type-1 AGN is mostly a matter of their activity level and the contrast of
their nuclear light to their host galaxy.

\begin{figure*}
\centering
\includegraphics[clip,angle=270,width=15cm]{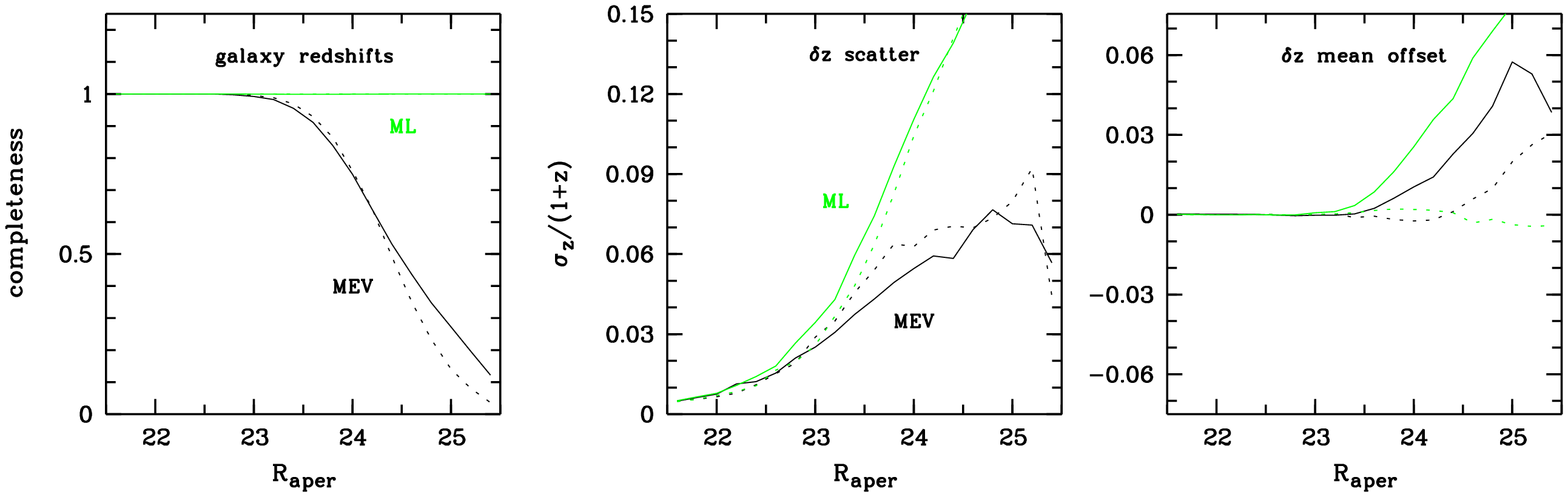}
\includegraphics[clip,angle=270,width=15cm]{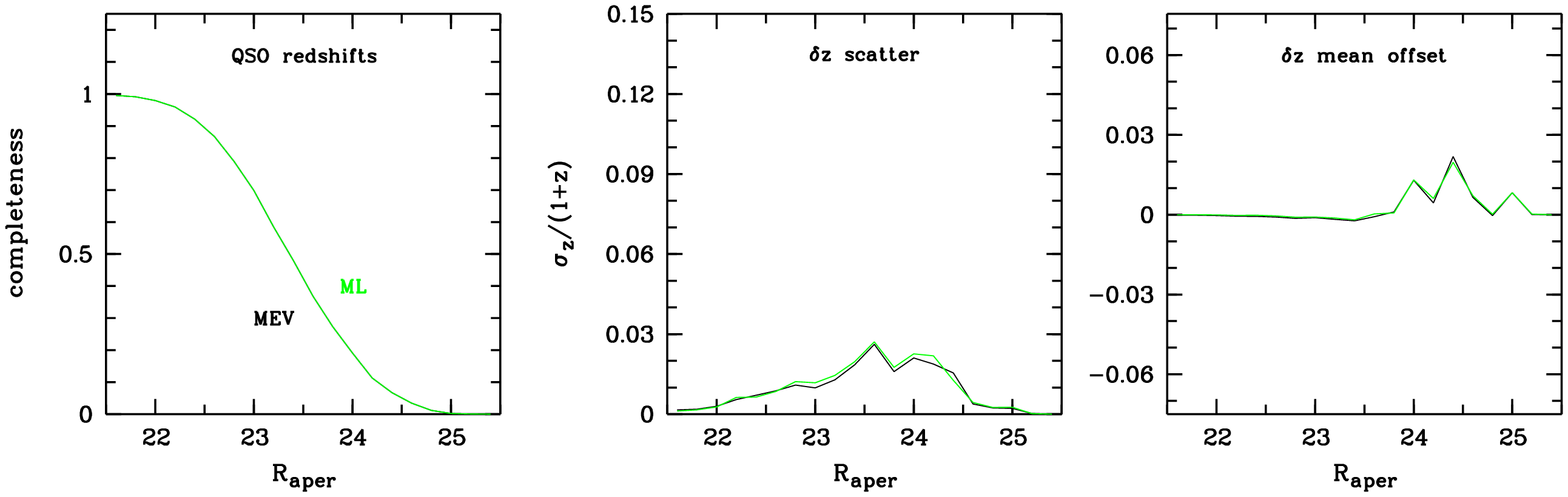}
\caption{Redshift quality expected from Monte-Carlo simulations: 
Galaxies at $z=0.1\ldots 1.1$ (top row, solid line: early types, dashed 
line: late types) and QSOs at $z=1\ldots 4$ (bottom row).
{\it Left panels:} Completeness of classification combined with redshift 
determination, for redshift values of the Minimum Error Variance (MEV, 
black line) and the Maximum Likelihood (ML, grey line) method.
{\it Center panels:} 1$\sigma$-scatter in true redshift error.
{\it Right panels:} mean offset in estimated $z$ from simulated $z$.
The QSOs appear to have extremely high redshift accuracy still at faintest
levels, but only because the completeness drops drastically and is reduced
to the most conspicuous QSOs. 
\label{p3_MoCa}}
\end{figure*}

\subsection{Expected redshift errors}

From the simulations we can estimate expected redshift errors by comparing
the input redshifts with recovered estimates. If objects are bright enough, 
redshifts are measured equally well for galaxies and QSOs, both of which 
have enough spectral features. If objects are faint, weak spectral features 
are washed out more effectively by noise than strong features, and it is 
not a-priori clear how the redshifts of different objects are affected.

When changing redshift, the fixed wavelength resolution of the filter set 
translates into an error that scales with $(1+z)$. Therefore, we shall 
discuss redshift errors only in terms of $\delta_z/(1+z)$ to remove this 
obvious dependency. 

When changing magnitude, we implicitly change the average redshifts and 
spectral types of a sample. Having eliminated the influence of redshift
above, we can split the sample into different spectral types and look only 
at the influence of magnitude. Essentially, a magnitude change is a change 
in S/N ratio across all filter fluxes and hence on the measured SED. To 
first order, redshift errors should be proportional to the radii of error 
ellipsoids in colour space, provided the mapping from redshift to colour 
space is not too non-linear. Because photometric errors increase inversely 
proportional to a decreasing flux, we expect mean redshift errors to scale 
as $\sigma_z/(1+z) \propto 1/F$ with $F$ being the mean flux of an object. 

However, Poisson noise is not the only source of photometric error because
the relative calibration of the passbands is also subject to uncertainties
as well as variations in real object spectra which are not reflected in the
sequence of templates. These additional effects cause mismatches between
object and templates which are non-Gaussian in shape and non-Poissonian in
magnitude dependence. They are actually independent of magnitude, although
the mean template mismatches could change due to a change in the underlying
sample. When going fainter, the calibration error should matter less and 
less in comparison to the Poisson error. But the faint object population is 
less well known and might harbour surprising challenges for the templates.

Furthermore, most known objects cover only a finite volume in colour space,
and hence require a finite error to be discriminated. At some low signal 
level, the classification will just break down, because there are plenty of
interpretations possible for the noisy observed SED. Here, some a-priori
knowledge from the literature could guide an interpretation if needed.

We like to repeat here our division of the magnitude scale with respect to 
the quality of classification and redshifts into three basic domains:

\begin{enumerate}
\item 
  the {\it quality saturation domain} at $R<22$ limited by systematic errors
\item
  the {\it quality transition domain} at $R=22\ldots 24.5$ limited by photon
  noise
\item
  the {\it quality breakdown domain} at $R>24.5$ which is useful only in
  conjunction with (and limited by) a-priori knowledge.
\end{enumerate}

\subsubsection{Galaxies}

We average the simulated redshift accuracy across a range in redshifts
chosen to match the bulk of the observed population, thus running from
$z=0.1$ to $z=1.1$. We split the full galaxy sample into red-sequence 
galaxies and star-forming blue-cloud galaxies to check for differences
in accuracy. In each magnitude bin, we determined the mean offset
among the redshift deviations $\delta_z/(1+z)$ to check for a possible 
bias, and calculated the scatter. Between the two galaxy types we found
little difference (see Fig.~\ref{p3_MoCa}). We also compared the redshifts 
from a Maximum-Likelihood (ML) estimator with the Minimum-Error-Variance
(MEV) estimator that we use routinely. As explained above, we ignore the
MEV measurement when its error goes above a threshold (see Sect.~4.5).
We found the scatter of MEV redshift deviations to be below the scatter 
of ML redshifts. At bright magnitudes, where both redshift exist for all
objects, this shows the superior performance of the MEV estimator (which
is by design). At faint magnitudes, only those objects have MEV redshifts
which have not too wide $p(z)$ distributions and are not expected to have
strong deviations in the first place. This selection keeps the scatter of
MEV redshift deviations much below the ML curve.

The simulations suggest that MEV redshifts should be roughly wrong by 
$\sigma_z/(1+z)\la 0.01$ at $R<22$, and throughout the quality saturation 
domain. Towards $R=24$ the mean redshift scatter should increase to 0.06
whereas the fraction of galaxies with MEV estimates has dropped to $\sim
75$\%. In the quality breakdown domain, the scatter does not increase by
much, but the fraction of objects with MEV estimates is extremely small.

\begin{figure*}
\centering
\includegraphics[clip,angle=270,width=0.8\hsize]{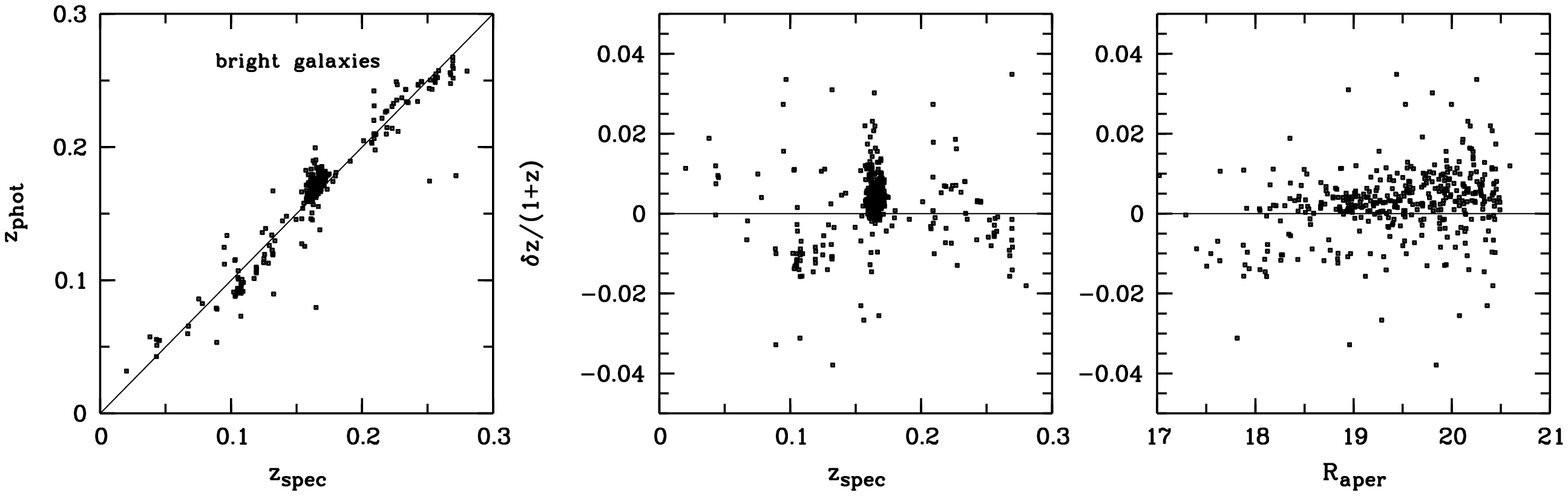}
\includegraphics[clip,angle=270,width=0.8\hsize]{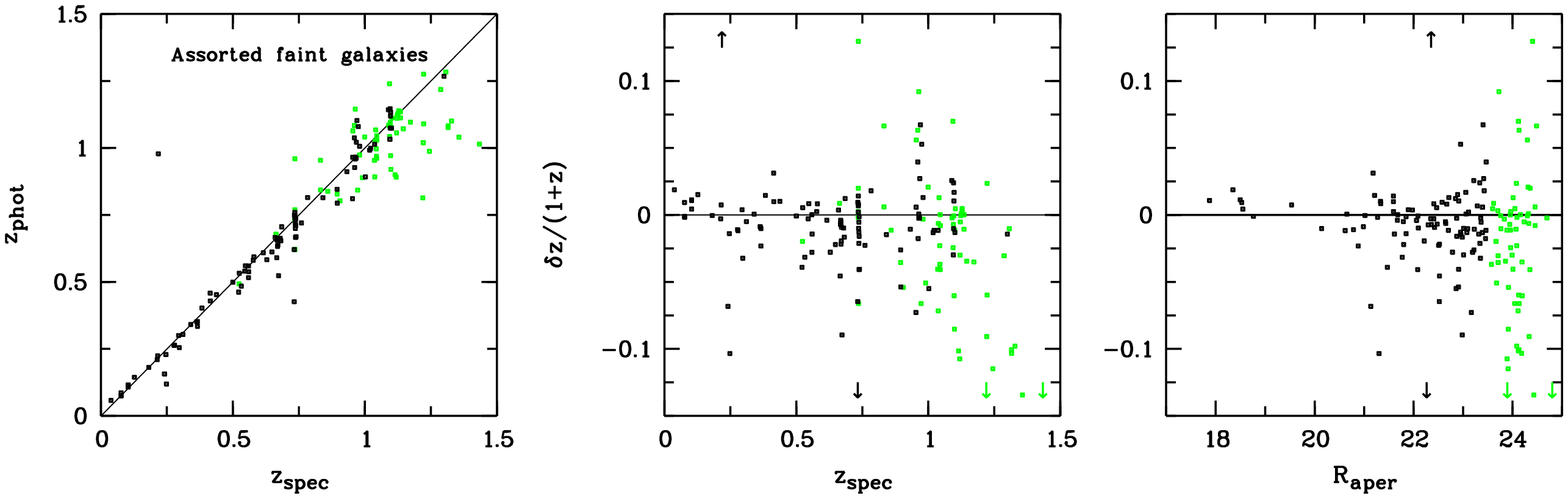}
\includegraphics[clip,angle=270,width=0.8\hsize]{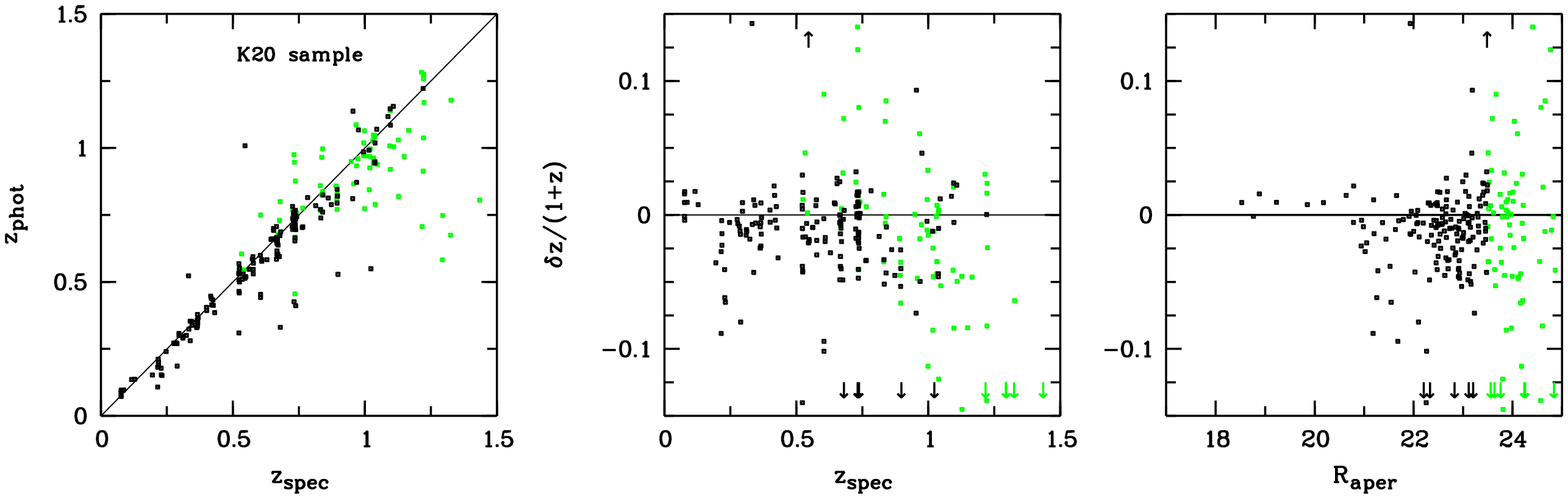}
\caption{Redshift quality of galaxies: 
{\it Top row:} 404 bright galaxies observed with 2dF, mostly containing 
galaxies from the supercluster field A901/902.
{\it Left panel:} Redshift comparison $z_{phot}$ vs. $z_{spec}$. 
{\it Center panel:} Redshift error $\delta z/(1+z)$ vs. redshift $z_{spec}$.
{\it Right panel:} Redshift error $\delta z/(1+z)$ vs. aperture $R$-band 
magnitude. The 1-sigma redshift error is $<0.01$, and the outlier rate for 
errors above 0.05 is $<$1\% (3 out of 404 objects with errors around 0.06).
{\it Center row:} A sample of 162 faint galaxies. Objects fainter than 
$R=23.5$ are shown in grey. Objects with redshift deviations above 0.15 are
shown as arrows. 
{\it Bottom row:} 247 galaxies from the K20 survey by Cimatti et al. Arrows 
in center or right column panels represent outliers with more than 15\% 
redshift error in (1+z).
\label{p3_gals}}
\end{figure*}

\begin{figure*}
\centering
\includegraphics[clip,angle=270,width=0.8\hsize]{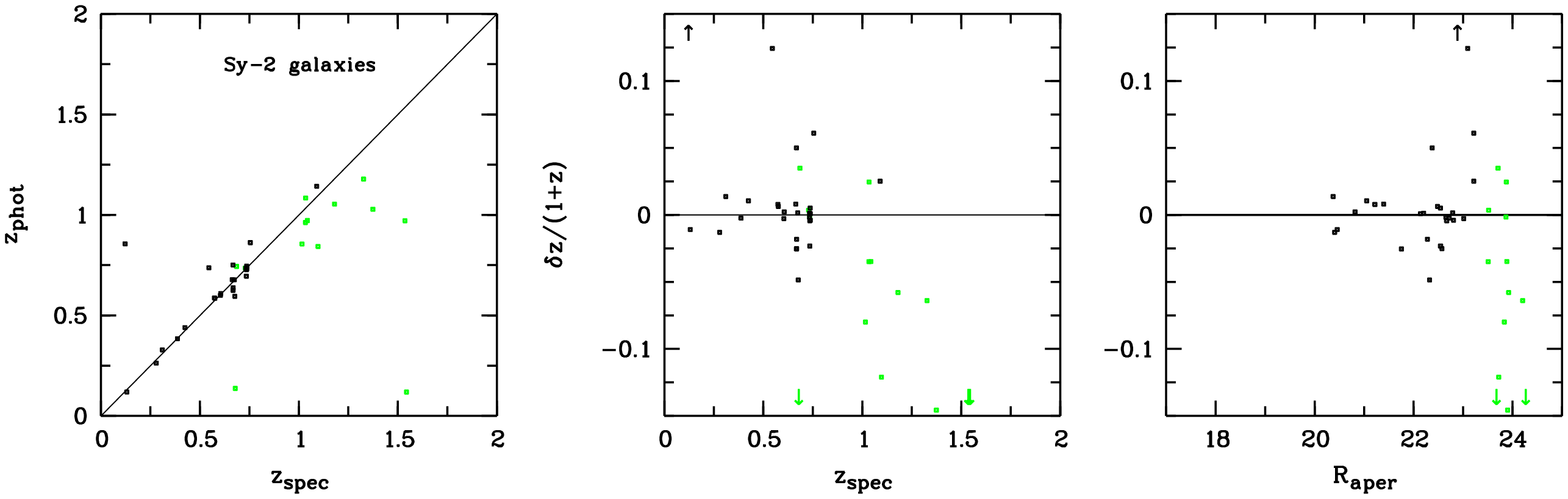}
\includegraphics[clip,angle=270,width=0.8\hsize]{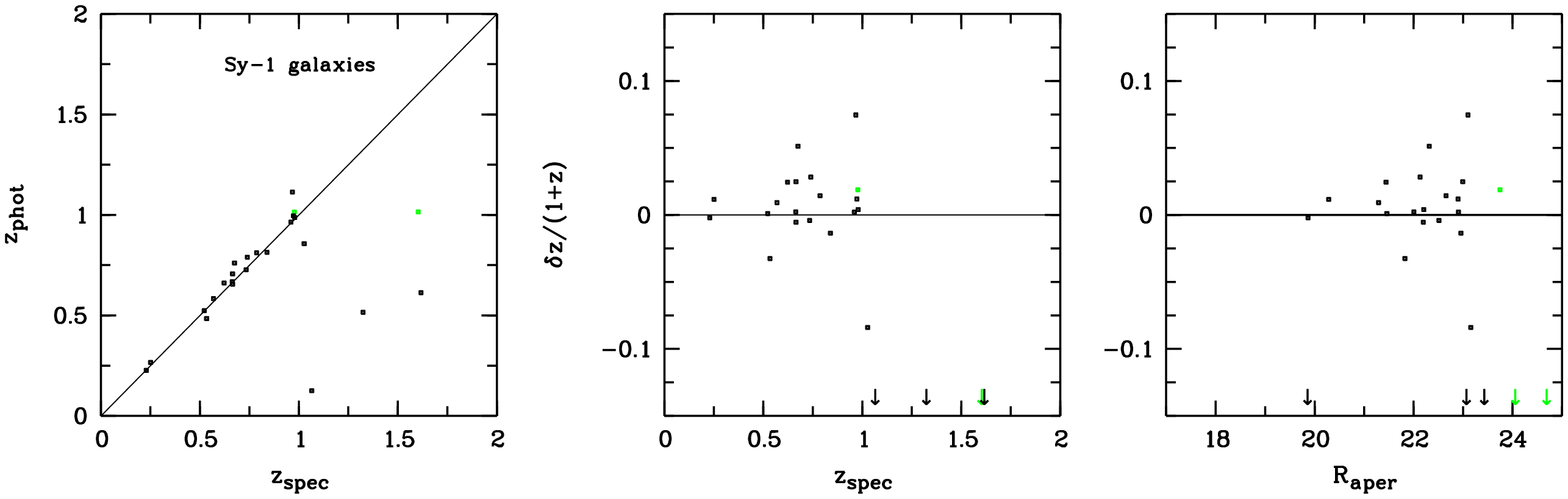}
\includegraphics[clip,angle=270,width=0.8\hsize]{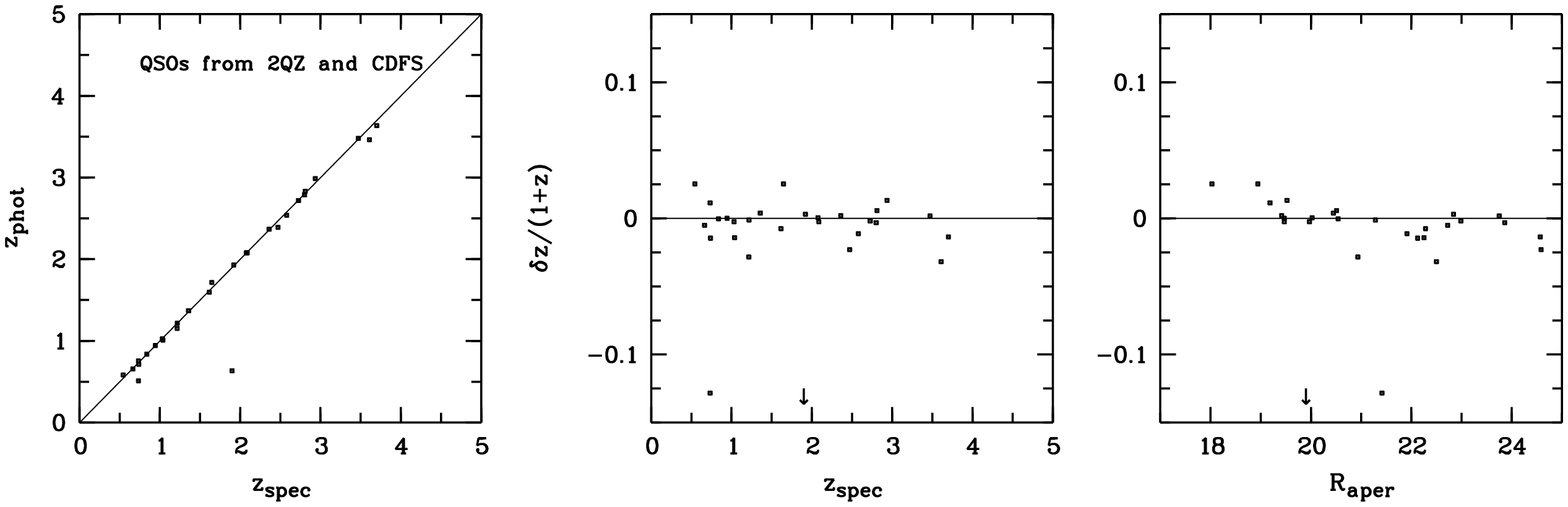}
\caption{Redshift quality of AGN: Panels are as in Fig.~\ref{p3_gals}. The 
active galaxy sample is from the Szokoly et al. (2004) follow-up of X-ray 
sources, and contains 24 Sy-1 and 41 Sy-2 objects, cut off here at $R<24$.
Their QSO sample contains 20 QSO which we have supplemented with seven QSOs 
from 2QZ on the S11 field and a QSO from slitless spectroscopy in the CDFS. 
The average redshift error of QSOs is $\approx 0.015$ with two outliers in 
27 objects. 
\label{p3_AGN}}
\end{figure*}

\begin{figure}
\centering
\includegraphics[clip,angle=270,width=\hsize]{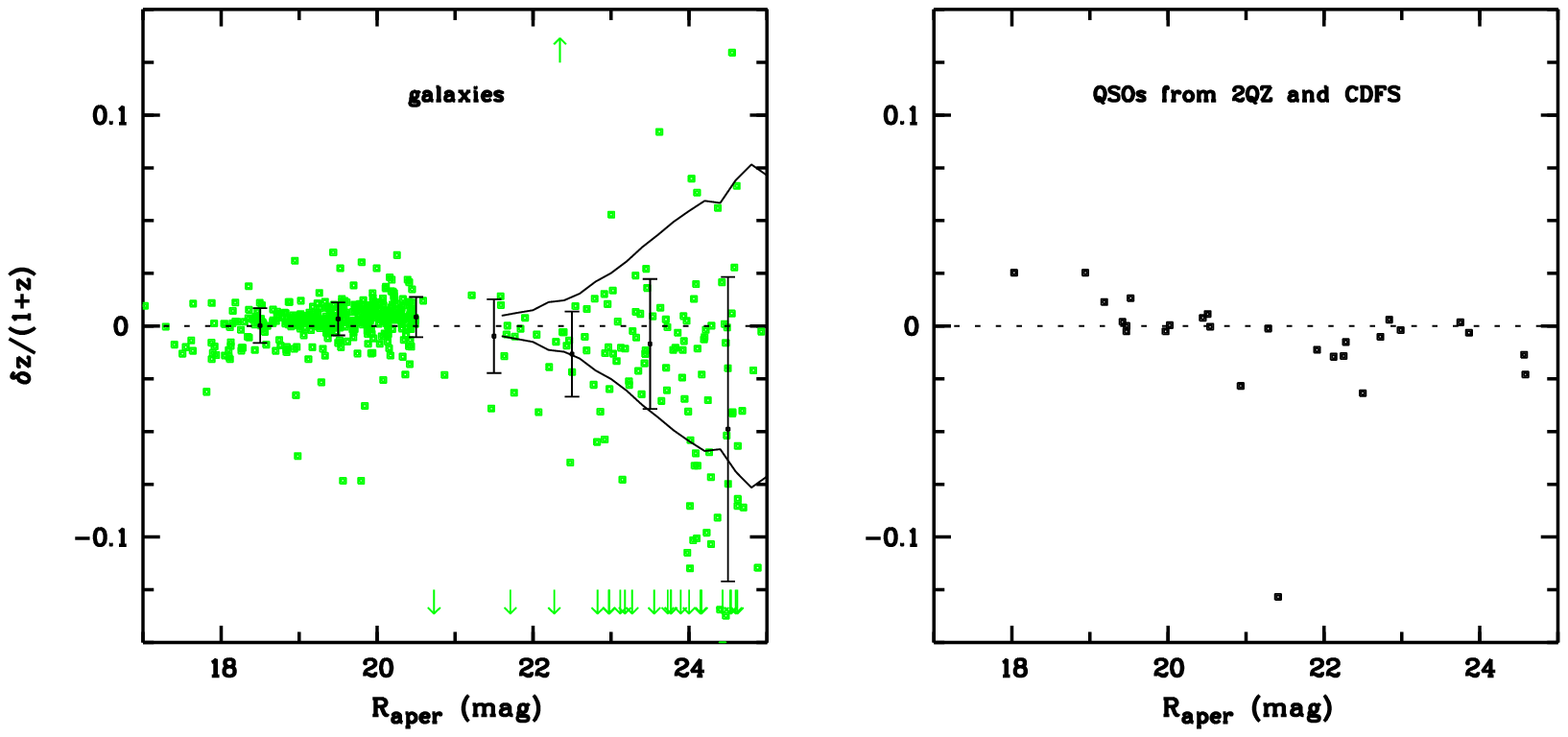}
\caption{Summary of galaxy redshift quality: This figure contains all 813 
galaxies from all available samples. Individual redshift deviations have 
been averaged within full magnitude bins, and the resulting zeropoint 
offsets and the 1-$\sigma$ scatters are shown as black error bars. 
Furthermore, the 1-$\sigma$ redshift errors predicted by the simulations 
are plotted as solid lines at $R>21.6$.
\label{allgal_dzs}}
\end{figure}

\subsubsection{Quasars}

We also determined the mean offset and scatter for quasar redshifts,
across the redshift range of the bulk sample, from $z=1$ to $z=4$. The
results for the ML and MEV estimator are virtually identical. Even the
completeness of the ML estimator drops with magnitude, because not all
quasars are identified as such. At faint levels, many simulated quasars 
are mistakenly identified as galaxies and have wrong redshifts assigned.
In Fig.~\ref{p3_MoCa} we only look at redshift deviations of recognized
quasars because only those would be in an observed quasar sample. We 
found the scatter and offset to be generally very low at all magnitudes.
This is because, whenever a quasar is recognized as such, its spectral
features were strong enough to get its redshift right, with an average 
error of $\sigma_z/(1+z)\la 0.015$.

\section{Spectroscopic performance test}

Currently, we have several independent spectroscopic samples available 
to test the quality of classification and redshifts in COMBO-17: there
are complete samples from 2dF spectroscopy on three COMBO-17 fields, 
X-ray follow-up in the CDFS (Szokoly et al. 2004), and the K20 survey 
(Cimatti et al. 2002)in the CDFS. The 2QZ survey overlapping with the 
S11 field is basically complete, but does not cover QSOs at all redshifts. 
An important but incomplete sample is the VLT/FORS spectroscopy on the 
CDFS started by GOODS. Altogether, these samples comprise spectra of 37 
stars, 813 inactive galaxies, 28 QSOs and 65 active galaxies. Almost all 
these objects are brighter than $R=24.5$.

\subsection{Stars}

The spectroscopic samples contain 37 stars. Four of these were observed
in Chandra follow-up and 33 mostly red stars were contained in the K20 
survey. The faintest one ($R=23.3$) out of the 37 was wrongly classified 
by us as a galaxy, while the other 36 were correctly identified as stars. 
We believe, that at $R\ga 23$ there could be problems with mixing up red 
stars and red-sequence galaxies. But at $R\la 23$ these findings suggest
that the stellar sample is complete.

\subsection{Galaxies}

\subsubsection{Bright galaxies observed with 2dF}

Spectroscopy of bright galaxies was available through the 2dFGRS survey 
on the S11 field with 39 galaxies including the cluster A1364 at $z\sim 
0.11$ and 14 galaxies on the CDFS field, most at $R\la 18$. Proprietary 
observations on the A901 field provided a larger and deeper sample of 
351 galaxies at $R\la 20$ including the cluster A901/902 at $z\sim 0.16$.
The total number of independent redshifts is 404. We will assume here 
that the classification quality is the same across all COMBO-17 fields. 
Fig.~\ref{p3_gals}, top row, displays the redshift quality of these 
bright galaxies. We find that 77\% of all galaxies have redshift errors 
$\delta_z/(1+z)$ below 0.01 and three objects, i.e. less than 1\%, deviate 
by more than 5-$\sigma$ from the true redshift.

\subsubsection{Assorted faint galaxy samples}

In the faint domain we have no very large and complete spectroscopic
sample available. Hence, we have opted for collecting assorted samples
from various authors as far as possible. Here, we include 31 inactive 
galaxies observed by Szokoly et al. (2004), 21 red-sequence galaxies
around $z\sim 1$ observed by Franx et al. (2004), and 110 galaxies 
observed with FORS2 by GOODS. The GOODS sample is larger
than this, but we excluded repeat measurements with the other sources
after confirming their consistency. At this stage, we restricted the 
sample to objects with secure line identification (Kuntschner, priv. 
comm.). More objects might become available soon. The top row of 
Fig.~\ref{p3_gals} shows the redshift quality of this assorted sample.

\begin{figure}
\centering
\includegraphics[clip,angle=270,width=\hsize]{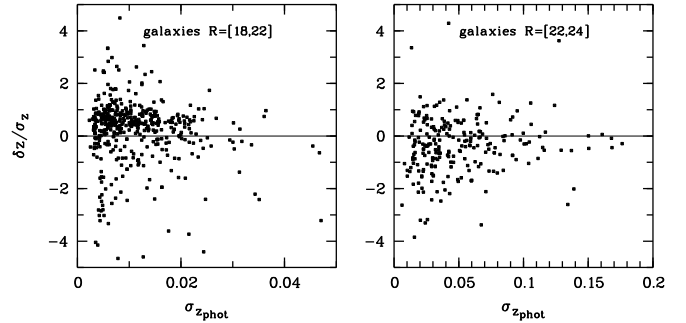}
\caption{Galaxy redshift errors: true deviations divided by estimated
errors vs. estimated errors. In a truely Gaussian scatter we expect 
$\sim 68$\% of a sample to fall into $[-1,+1]$. In our sample this is
in fact the average behaviour although there is a non-Gaussian tail to
larger-than-expected deviations.
\label{dzgs}}
\end{figure}

\subsubsection{K20 sample}

We present the redshift comparison with the K20 sample (Cimatti et al. 2002)
separately, because it is a complete sample, at least in terms of $K$-band 
selection, although not in terms of an $R<24$ selection. It preferentially 
contains red objects, while faint blue galaxies would be underrepresented. 
Excluding stars and broad-line AGN we selected 247 objects with $K_s<20$ and
their spectroscopic quality equal to 1, signifying reliable redshift. The 
COMBO-17 redshift quality from this comparison is shown in the bottom row of 
Fig.~\ref{p3_gals}. We see the same result as for the assorted sample above 
except for a somewhat higher outlier rate.

\subsubsection{Galaxy summary}

At $R\la 24$ and $z\la 1$ we seem to have (1-$\sigma$) redshift errors that 
could be described analytically by
\begin{equation}
   \sigma_z/(1+z) \approx 0.007 \times \sqrt{1+10^{0.8 (R-21.6)}}  ~ .
\end{equation}

The rate of outliers appears to be below 5\% at $z\la 1$. We would like to 
note, that even spectroscopic redshift determination can be challenging if
only one line is clearly visible. In such cases, the overall SED shape from
COMBO-17 or its own redshift estimate can be helpful to spectroscopic work
in order to reach a final interpretation of an observed line. Finally, any
blended objects will completely fool the COMBO-17 approach which probably 
can not decompose a spectrum into two. Spectroscopy of sufficient quality 
and spatial resolution can deal with those rare challenges much better. 

In Fig.~\ref{dzgs} we compare the true redshift deviations with individual 
1-$\sigma$ redshift uncertainties estimated by the template matching code. 
As expected we find a scattered distribution of true deviations around the 
expected mean. We find, that objects with larger estimated errors have true 
deviations as expected from the mean or slightly below. In contrast, objects
with extremely small estimated errors scatter a bit more than expected from 
the error estimation. A Gaussian distribution puts $\sim$68\% of a sample 
below its 1-$\sigma$ cut. We find numbers between 53\% and 78\% in different
subsamples of our galaxies. Bright, large-error galaxies ($R=[18,22]$ and
$\sigma_z>0.01$) have more accurate redshifts than expected: 78\% are at
$\delta z < \sigma_z$. Faint, small-error galaxies ($R=[22,24]$ and 
$\sigma_z<0.03$) have less accurate redshifts than expected: 53\% are at
$\delta z < \sigma_z$. Bright, small-error objects and faint, large-errors
objects behave as expected with $\sim70$\% at $\delta z < \sigma_z$.

\begin{table}
\caption{COMBO-17 classification and 2dF spectroscopy of QSO candidates 
from the 2QZ survey on the S11 field. Note, that COMBO-17 detected all
true quasars at the right redshift, and that it was able to discard all
false candidates except for a white-dwarf/M-dwarf binary, for which no
matching templates existed. \label{2QZ}}
\begin{tabular}{lclc|cl}
\hline \noalign{\smallskip} \hline \noalign{\smallskip} 
\multicolumn{4}{c}{COMBO-17}        & \multicolumn{2}{c}{2QZ} \\
Identifier & $R$ (mag) & class      & $z_{phot}$ & $z_{2dF}$ & class \\
\noalign{\smallskip} \hline \noalign{\smallskip}
S11-03978  & 19.47 & QSO            &    2.078      & 2.086 & QSO    \\
S11-17931  & 17.99 & Star           &               &       & cont   \\
S11-18345  & 20.00 & QSO            &    2.078      & 2.077 & QSO    \\
S11-19860  & 19.55 & Galaxy         &    0.134      &       & ??     \\
S11-22644  & 18.89 & WDwarf         &               &       & star   \\
S11-36451  & 19.53 & QSO            &    2.987      & 2.935 & QSO    \\
S11-36868  & 19.57 & QSO            &    0.596      &       & WD(DA) \\
S11-40157  & 20.45 & QSO            &    1.367      & 1.358 & QSO    \\
S11-40953  & 19.44 & QSO            &    0.944      & 0.944 & QSO    \\
S11-42872  & 18.01 & QSO            &    1.714      & 1.647 & QSO    \\
S11-43560  & 19.43 & QSO            &    2.367      & 2.360 & QSO    \\
S11-45585  & 20.07 & Star           &               &       & star   \\
S11-48688  & 20.09 & Star           &               &       & ??     \\
S11-52614  & 20.51 & WDwarf         &               &       & ??     \\
\noalign{\smallskip} \hline
\end{tabular}
\end{table}

\subsection{AGN sample in CDFS and S11 field}

The spectroscopic follow-up of X-ray sources in the CDFS \cite{Szo04}
provides a very good sample to test the completeness of AGN detection
in COMBO-17 (see Fig.~\ref{p3_AGN}). As already discussed above, the
{\it QSO} class is a complete sample of type-1 AGN at $M_B<-21.7$ and
contains an incomplete set of low-luminosity type-1 AGN. Whenever the
host galaxy dominates the SED and nuclear light from a low-luminosity 
AGN is only a fraction of that, COMBO-17 considers the object as a 
{\it Galaxy} and basically determines the redshift of the host galaxy.
From COMBO-17 alone, we have no means to identify the nuclear activity 
in AGN of such low luminosity. The same applies to type-2 AGN, where
again the host galaxy dominates the visual SED by far.

Another question is whether the host galaxy redshifts of low-luminosity 
AGN are estimated correctly, or whether the presence of the AGN light 
leads to problems. Hence, we investigate the redshift quality of that
fraction of the COMBO-17 galaxy sample, which has been classified as 
Sy-1 or Sy-2 by Szokoly et al. (2004) on the basis of X-ray data and
optical spectroscopy.

We find that for objects at $z\la1$ the redshift estimation behaves
like it does for normal galaxies with the exception of a remarkably 
faint dwarf Sy-2 object (see also Sect.~4.6). This object has $R_{total}
=22.6$ and shows an SED that perfectly matches a red-sequence galaxy at 
$z\approx 0.85$. But the spectroscopy shows emission lines at $z=0.122$
on top of a remarkably red continuum. If all the light originated in
fact from a source at $z=0.122$, this galaxy would be a dwarf Sy-2 with 
a total luminosity of $M_B\approx -14.5$ and a truely unusual continuum.
If the object was a blend of two and most of the light came from a 
background galaxy, the host of this AGN would be even fainter.

At $z>1$ we find that Sy-2 galaxies behave still much the same as normal
galaxies, but Sy-1 galaxies produce three outliers. Here we are probably 
confronted with restframe UV SEDs which are a mix of host galaxy light 
and nuclear light confusing the classification. The Sy-1 galaxies are the
weakest feature of classifier. Presently it is not clear, whether sufficient 
templates could be defined which could alleviate this problem.

Finally, the redshifts of QSOs (including seven objects from 2QZ and one
object from slitless spectroscopy on the CDFS) have 
errors of $\sigma_z/(1+z) \approx 0.015$, independent of magnitude. This 
is plausible because QSOs have sufficiently strong spectral features to 
pin down their redshifts, except for a low-redshift outlier, which is off 
by 0.13, and a $z\sim 2$ outlier whose redshift is completely wrong. 
Providing identifications and photometric redshifts for quasars at 1.5\% 
accuracy is one of the strongest advantages of COMBO-17.

We would like to report also on spectroscopic cross check available from
the 2dF Quasar Redshift Survey (hereafter 2QZ, see Tab.~\ref{2QZ}) on the
S11 field of COMBO-17. The seven QSOs identified there by 2QZ have already
been included into Fig.~\ref{p3_AGN}. But in addition, we present the full
table of 2QZ candidates and their spectroscopic ID as obtained with 2dF in
Tab.~\ref{2QZ}. We compare the spectroscopic result with our classification 
and demonstrate that COMBO-17 provides almost equivalent information, with
two restrictions: (i) COMBO-17 misinterpreted an M-dwarf/white-dwarf binary
as a quasar; (ii) The COMBO-17 redshifts are obviously less accurate at a
level of $\sigma_v \la 5000$~km/sec.

\section{Summary}

Wolf, Meisenheimer \& R\"oser (2001) showed that medium-band surveys deliver 
more accurate object classifications and photometric redshifts than broad-band 
surveys, while not consuming more telescope time. As a result, the COMBO-17 
survey was started to obtain a large redshift catalogue of galaxies and AGN 
for evolutionary population studies, including weak lensing observations.

In this paper, we have discussed in detail the quality of the classification
and photometric redshifts of galaxies and quasars in COMBO-17. We have shown
that the identification of stars is complete to $R\la 23$ (deeper for M~stars).
We have demonstrated that the identification of type-1 AGN is complete above
luminosities of $M_B=-21.7$ at all redshifts from 0.5 to 5. The photometric
redshifts of galaxies in COMBO-17 are better than 0.01 at bright magnitudes
($R<21$) and increase with photometric noise to $\delta_z/(1+z) \approx 0.06$
at $R=24$. Fainter than $R=24$, COMBO-17 is not particularly useful because
the medium-bands are too shallow then. We have demonstrated that we routinely
obtain photometric redshifts of quasars and luminous Seyfert-1 galaxies to an
accuracy of $\delta_z/(1+z) \approx 0.015$. 

We have demonstrated that the medium-band approach indeed delivers the 
expected performance which motivated the survey COMBO-17. In this paper 
we now deliver the COMBO-17 database from one particular patch of sky to 
the community for public exploitation. The published database includes 
images and a catalogue with 63,501 objects. Classification and redshifts 
are typically reliable at $R<24$, where we find $\sim 100$ quasars, $\sim 
1000$ stars and $\sim 10000$ galaxies.

We included the Chandra Deep Field South from the very beginning in the 
COMBO-17 project. A multitude of deep observations was expected across a wide 
range of photon energies, creating a unique data set for studies of galaxy 
evolution. The COMBO-17 approach has allowed us to get hold of a large 
redshift catalogue on the field. This includes the implicit selection of 
$\sim 100$ luminous type-1 AGN with photometric redshifts as accurate as 
$\sim 5000$~km/sec. We are now in a position to make this catalogue available 
for general use and hope to feed many more dedicated follow-up studies.

\begin{acknowledgements}
C. W. was supported by the PPARC rolling grant in Observational 
Cosmology at University of Oxford. M. K. was supported by the 
DFG--SFB 439 at University of Heidelberg. E. F. B. was supported
by the European Community's Human Potential Program under contract
HPRN-CT-2002-00316, SISCO. 

\end{acknowledgements}

\end{document}